\documentclass{article}

\usepackage{arxiv}

\usepackage[utf8]{inputenc} 
\usepackage[T1]{fontenc}    

\usepackage{amsmath,amssymb,amsfonts}
\usepackage{graphicx}
\usepackage{textcomp}
\usepackage{xcolor}
\usepackage{booktabs} 
\usepackage{algorithm}
\usepackage[noend]{algpseudocode}
\usepackage{url}
\usepackage{color}
\newcommand\eat[1]{}

\newtheorem{definition}{Definition}

\title{Continuous Outlier Mining of Streaming Data in Flink}

\author{
 Theodoros Toliopoulos, Anastasios Gounaris, Kostas Tsichlas, Apostolos Papadopoulos \\
Department of Informatics, Aristotle University of Thessaloniki, Greece\\ \{tatoliop,gounaria,tsichlas,papadopo\}@csd.auth.gr
 \AND
 Sandra Sampaio \\
 The University of Manchester, UK \\           
 s.sampaio@manchester.ac.uk
}

\begin{document}
\maketitle

\begin{abstract}
In this work, we focus on \textit{distance-based outliers} in a metric space, where the status of an entity
as to whether it is an outlier is based on the number of other entities in its neighborhood. In  recent years, several solutions have tackled the problem of distance-based outliers in data streams, where outliers must be mined continuously as new elements become available. An interesting research problem is to combine the streaming environment with massively parallel systems to provide scalable stream-based algorithms.
However, none of the previously proposed techniques refer to a massively parallel setting. Our proposal fills this gap and investigates the challenges in transferring state-of-the-art techniques to Apache Flink, a modern platform for intensive streaming analytics.
 We thoroughly present the technical challenges encountered and the alternatives that may be applied. We show speed-ups of up to 117 (resp. 2076) times over a naive  parallel (resp. non-parallel) solution in Flink, by using just an ordinary four-core machine and a real-world dataset. When moving to a three-machine cluster, due to less contention, we manage to achieve both better scalability in terms of the window slide size and the data dimensionality, and even higher speed-ups, e.g., by a factor of 510. Overall, our results demonstrate that oulier mining can be achieved in an efficient and scalable manner. The resulting techniques have been made publicly available as open-source software.
\end{abstract}

\keywords{
distance-based outlier detection \and Flink \and data streams}

\section{Introduction}

Outlier analysis forms a key mechanism in modern data science and analytics~\cite{Aggarwal2013}, aiming to detect objects that, as defined in \cite{Johnson92}, appear to be inconsistent with the remainder of the objects in the same dataset. They are used in a variety of applications, such as fraud detection, spam detection and medical diagnosis, just to name a few. One of the most commonly used definitions
for an outlier is the distance-based one \cite{KNT00}, where an object is considered an outlier if it does not have more than $k$ neighbors within a distance up to $R$.  Continuous outlier detection in data streams deals with the problem of keeping an updated list of all outliers after each new object arrives and/or expires. Apparently, when data grows large, this becomes a  challenging task, since applying even an one-pass algorithm to all active data is prohibitively expensive. To improve efficiency and scalability, the main target of this work is to propose massively parallel solutions for continuous outlier detection in data streams.

Briefly, the relevant state-of-the-art falls into two categories. The first category contains efficient non-par-allel solutions for streaming outlier detection, e.g., the works described in \cite{KontakiGPTM16,YRW09,cao2014scalable,AF07}. The second category contains parallel solutions for outlier detection, where, to date, there is a single proposal that assumes modern distributed computing platforms, such as MapReduce  \cite{CYK+17}; nevertheless, this solution does not deal with streaming data. The novelty of our proposal, to the best of our knowledge,  is that it proposes, for the first time, the combination of  both massive parallelism and continuous distance-based outlier detection in a streaming setting. Orthogonally, outlier detection techniques are typically classified as either exact or approximate. We target exact solutions in this work, and more specifically, we aim to transfer the main features of the state-of-the-art non-parallel exact solutions, which are investigated in \cite{TranFS16}, to a parallel setting in an efficient manner.

Devising efficient parallel solutions for this problem involves addressing a series of important issues. First, outlier detection algorithms in data streams involve windows that cannot be partitioned into non-overlapping partitions, among which no communication is required. Second, low latency is of high significance in order to deliver results in a timely manner. Third, state information needs to be kept between window slides in order to avoid unnecessary recomputations.  We provide solutions to the above issues through the application of key ideas obtained from existing non-parallel techniques, such as the ones described in  \cite{AF07,KontakiGPTM16} to the Flink\footnote{\url{https://flink.apache.org/}} platform.

This work\footnote{This is an extension to our paper titled ``Parallel Continuous Outlier Mining~in Streaming Data" presented in 5th IEEE International Conference on Data Science and Advanced Analytics (DSAA), 2018 \cite{DSAA18}.} aspires to become a reference point for all future work on streaming outlier detection in massively parallel settings and its main contributions are summarized as follows:

\begin{enumerate}
\item[(i)]
We explore a series of implementation alternatives, differing in the algorithmic features they employ and in the way data is partitioned. Compared to the earlier work described in \cite{DSAA18}, we present a novel Vantage-Point (VP) tree-based  value partitioning technique that is suitable for arbitrary metric spaces, as well as the encapsulation of the notion of window slicing.
\item[(ii)]
We provide thorough experimental evaluation results. Compared to the earlier work described in \cite{DSAA18}, we additionally investigate the behavior of continuous outlier detection algorithms running on a cluster consisting of multiple multi-core physical machines, and we assess  the impact of the use of priority queues and M-tree, employed in state-of-the-art non-parallel solutions, such as the ones described in \cite{TranFS16}.
\item[(iii)]
We offer the source code as an open-source library.\footnote{The code repository may be accessed at \url{https://github.com/tatoliop/
parallel-streaming-outlier-detection}}
\end{enumerate}
In summary, we show that our best performing alternative, when tested on a real-world dataset, can yield speed-ups of up to 117 (resp. 2076) times over a naive parallel (resp. non-parallel) solution
in Flink using just a commodity four-core machine. Similar performance is observed for synthetically generated datasets as well.  Due to resource contention, we observe that, on a single machine, the scalability is not good regarding the data dimensionality and size of the window slide. Both problems are tackled in a cluster environment, where even higher speed-ups are observed, e.g., 510 times faster using three machines.

The remainder of this work is structured as follows.  Section~\ref{sec:back} contains background material on parallel streaming platforms
and outlier detection algorithms. Section~\ref{sec:simple} introduces our first parallel solution, which is extended in Sections~\ref{sec:advanced} and \ref{sec:pmcod}. Performance evaluation results are offered in Section~\ref{sec:eval}. We close this work with a discussion of related work and issues relating to extensions to our techniques in Sections ~\ref{sec:rw} and ~\ref{sec:disc}, respectively.

\section{Fundamental Concepts and Background}
\label{sec:back}

The purpose of this section is to make the article as self-contained as possible.
We split background material into two parts, referring to the main massively parallel platform alternatives for streaming data and the distance-based outlier algorithms that inspired our solutions, respectively. 

\subsection{Parallel Frameworks and Streaming Semantics}
\label{sec:frameworks}

In this section, the massively parallel platforms for streaming data that have been considered are presented in order to explain the reasons why Flink has been selected. In addition, the main streaming semantics of windows and timestamps are explained.

\subsubsection{Parallel Frameworks for Streaming Applications}
Three of the main parallel streaming platforms were examined, before choosing the most suitable one for the purposes of this work, namely: (i) Storm\footnote{\url{https://storm.apache.org/}}, (ii) Spark\footnote{\url{https://spark.apache.org/}} and (iii) Flink. 

Apache Storm is the first widely used large scale stream processing framework. Storm is able to connect with a number of queuing systems, such as Kestrel, Kafka and Amazon Kinesis, and  any database system. Storm provides low latency by using a record acknowledgment architecture, where  each record that is processed by an operator sends back an acknowledgment to the previous operator. Its architecture is fault-tolerant providing \emph{at-least-once} semantics. In case of  failure, the data is re-processed, but this may result in duplicate production. To offer \emph{exactly-once} semantics, there is a high level Storm API called Trident that  employs micro-batches.

Spark Streaming also enables scalable, high-through-put and fault tolerant processing of data streams. It supports many data sources such as Kafka, Flume, Twitter and TCP sockets. The processed data can be written to filesystems and databases, such as HDFS (Hadoop Distributed File System) and Cassandra. Spark divides the input streams into micro-batches. Each of these batches goes through a processing step generating another micro-batch (result stream) until it has passed through all steps and the final result is written into a data sink. Spark Streaming is fault-tolerant supporting \emph{exactly-once} semantics with a high-throughput by using the micro-batch architecture. This architecture, however, incurs  higher latency than the continuous streaming approach employed by Storm and Flink, because of the delay caused by the micro-batches.

Apache Flink is another massively parallel platform for continuous stream processing providing low-latency, high-throughput and fault tolerance. Flink supports a number of data connectors including Kafka, Amazon Kinesis and Twitter along with data sinks such as HDFS, Cassandra and ElasticSearch. Flink is a framework that provides \emph{exactly-once} semantics without resorting to micro-batches. Using a snapshot algorithm, it periodically generates state snapshots of a running stream topology, storing them in persistent storage, such as HDFS. In case of failure, Flink restores the latest snapshot from the storage and rewinds the stream source to the point where the last snapshot was taken. This approach combines the \emph{exactly-once} semantics with low latency stream processing. Flink also provides a high level API, facilitating the partitioning of a stream into windows and the development of processing operators.

In summary, Storm provides low latency, similar to Flink, but it incorporates the \emph{at-least-once} semantics, which allows duplicates to pass through the process in case of failures. Spark, like Storm Trident, provides \emph{exactly-once} semantics with the use of micro-batches. The main drawback of micro-batches is the higher latency compared to the continuous processing model of Storm and Flink. Finally, Flink combines the advantages of Storm and Spark, namely low latency regarding the continuous record processing, coupled with the \emph{exactly-once} semantics. In addition, Flink naturally supports both time- and count-based windows, which is not the case for Spark, since micro-batches essentially correspond to time-based sliding windows (see discussion below).

\subsubsection{Streaming Semantics}
A continuous stream is an infinite  sequence of data points. Each data point \textit{o} is annotated with its arrival time,  \textit{o.t}. The analysis of such infinite streaming data requires different techniques than those for finite datasets. A common approach is to adopt the notion of  \textit{window}, which refers to the most recent data items. Windows are typically small enough so that they can be stored in main memory, either of a single machine or of a parallel cluster.
Windowing essentially splits the data stream into either overlapping (\textit{sliding windows}) or non-overlapping (\textit{tumbling windows}) finite sets of data points. Orthogonally, the splitting can be based either on the time of arrival of the data points (\textit{time-based windows}) or on the number of data points (\textit{count-based windows}). In the former case, the window size $W$ is measured in time units, while in the latter case  the size corresponds to the number of the most recent data items held. In time-based windows,
\textit{W} is defined by the minimum and maximum timestamps for data items in order to be included in the window, denoted as \textit{W.start} and \textit{W.end}, respectively. More specifically, \textit{W=W.end-W.start}.

\begin{figure} [tb!]
	\centerline{\includegraphics[width=0.65\linewidth]{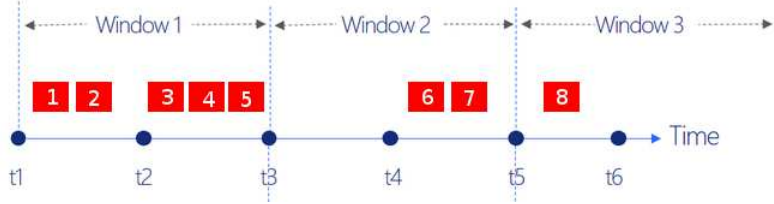}}
	\caption{Tumbling Windows}
	\label{fig:tumbling_window}
\end{figure}

 Figure \ref{fig:tumbling_window} shows a stream discretized in three windows based on time. The windows are non-overlapping with \textit{W} = 2 time units. 
 In a time-based window, the amount of data in the window varies through time and the contents of consecutive windows are disjoint sets. Tumbling windows conceptually divide a stream into non-overlapping partitions.

\begin{figure}[tb!]
	\centerline{\includegraphics[width=0.65\linewidth]{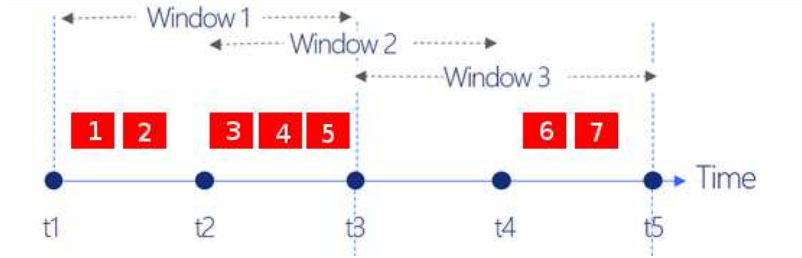}}
	\caption{Sliding Windows}
	\label{fig:sliding_window}
\end{figure}

  In this work, we focus on sliding windows, which generalize the tumbling ones, and our techniques support both time- and count-based windows.  Therefore, without any loss of generality, whenever we use the term window, we will be referring to a sliding time-based window.
 Figure \ref{fig:sliding_window} shows examples of such windows.
 In sliding windows, the magnitude of each slide is denoted as $S$.  Every time the window moves by \textit{S},  \textit{W.start} and \textit{W.end} are increased by $S$ as well. For example, in Figure \ref{fig:tumbling_window}, $S$=2 time units, and in Figure \ref{fig:sliding_window}, $S$=1 time unit. In each slide, some points may expire, i.e., they are dropped, while new points are included in the current window.
Table \ref{table:symbol_int} summarizes the notation used throughout the paper.


\subsection{Problem Definition and Non-Parallel Solutions}
\label{sec:main-algs}

The problem of continuous distance-based outlier detection is defined formally as follows.

\begin{definition}
  Given a set of objects $\mathbb{O}$ and the threshold parameters $R$ and $k$, for each window slide $S$, report all the objects $o_i$ for which the number of neighbors $o_i.nn < k$, i.e., the number of objects $o_j,~j \ne i$ for which $dist(o_i,o_j) \le R$ is less than $k$.
\end{definition}

The main challenges stem from the fact that all active objects need to be continuously assessed during their lifetime, since an object may change its status as many times as the number of slides during which it remains within the window.

We exclusively focus on exact solutions.
There are several exact algorithms for continuous outlier detection in data streams, such as exact-Storm \cite{AF07}, Abstract-C \cite{YRW09}, LUE \cite{KGP+11,KontakiGPTM16}, DUE \cite{KGP+11,KontakiGPTM16}, COD \cite{KGP+11,KontakiGPTM16}, MCOD \cite{KGP+11,KontakiGPTM16} and Thresh\_LEAP \cite{cao2014scalable}\footnote{Reference implementations are provided in the MOA tool \cite{KGP+13} \url{http://moa.cs.waikato.ac.nz/} and \url{https://infolab.usc.edu/Luan/Outlier/}}. All these algorithms assume a centralized environment. In a recent impartial comparison presented in \cite{TranFS16}, these algorithms were compared in terms of their memory consumption and CPU time. The comparisons were made using four data-sets with varying dimensionality and settings of window length \textit{W}, window slide \textit{S}, number of neighbors \textit{k} and neighbor range \textit{R}.
This study showed that MCOD is superior  across multiple datasets in most stream settings. In addition, Thresh\_LEAP and MCOD displayed the lowest memory consumption and CPU times, while exact-Storm, Abstract-C and DUE are the slowest and most memory consuming algorithms.
Based on these findings, MCOD has served as our preferred basis for our solution to the problem of parallelization of continuous distance-based outlier detection algorithms. However, in our investigation, we start with a simpler and easier to parallelize algorithm, namely exact-Storm, which contains several key elements in common with MCOD, such as index structures for range queries and safe inliers, and so it represents a preliminary step towards our proposed solution. We also use the time-slicing notion of the Thresh\_LEAP algorithm as one of the extensions.
The key details of these three algorithms are discussed in the following.

\begin{table}[tb!]	
	\centering
		\begin{tabular}{|p{0.21\columnwidth}|p{0.7\columnwidth}|}
			\hline
			{\bf Symbol} & {\bf Short description} \\ \hline
			\textit{W}         	& The size of the stream window         \\ \hline
			\textit{S}         	& The slide of the stream              	\\ \hline
			\textit{W.start}   	& The starting timestamp of the window 	\\ \hline
			\textit{W.end}   	& The ending timestamp of the window 	\\ \hline
			$\mathbb{O}$              & The set of  data objects (or points) in the stream                        			\\ \hline
			$o_i \in \mathbb{O} $              & The $i^{th}$ data object in the stream                        			\\ \hline
			\textit{o.id}            & The object identifier \textit{o} (either $i$ for $o_i$ or any other identifier)                           	\\ \hline
			\textit{o.value}            & The value of \textit{o}                           	\\ \hline
			\textit{o.t}            & The arrival time of  \textit{o}                           	\\ \hline
			\textit{o.count\_after} & The number of succeeding neighbors of \textit{o}             	\\ \hline
			\textit{o.nn\_before}   & A list with the arrival time of the preceding neighbors of \textit{o} 	\\ \hline
			\textit{o.nn}   & The count of neighbors of \textit{o} 	\\ \hline
			$\mathbb{PO}$  			& A list containing data points that are potentially outliers \\ \hline
			$R$ & The distance threshold in the outlier definition \\ \hline
			$k$ & The neighbor count threshold in the outlier definition \\ \hline			
			$dist(o_i,o_j)$ & The distance function between objects $o_i$ and $o_j$ \\ \hline	
			$P$ & set of Flink partitions, each handled by a separate Flink node  \\ \hline			
		\end{tabular}
\caption{Frequently used symbols and interpretation}
\label{table:symbol_int}
\end{table}

\subsubsection{Exact-Storm}
\label{sec:exact-storm}
Two of the key operations in distance-based outlier detection are the distance computation between objects and the continuous examination of the neighborhood of each object. To avoid a quadratic number of comparisons in each slide, appropriate indices are required. To this end, exact-Storm uses a data structure called ISB to store the data points in nodes. A node is a record containing the data point \textit{o}, the arrival time of the point \textit{o.t}, the number of succeeding neighbors \textit{o.count\_after} and a list \textit{o.nn\_before} of size \textit{k} containing the arrival time of the preceding neighbors of \textit{o} (i.e., each node contains a different data stream  object along with some metadata). This data structure is a pivot-based index that provides support for fast range query search in any metric space. The range query, given a data point \textit{o} and the range \textit{R}, returns the nodes in the ISB whose distance to \textit{o} is less than or equal to \textit{R}.

A sketch of the algorithm steps in each slide is as follows. For each new data point \textit{o}, a node is created as described above. Then, a range query is issued on the ISB structure to find the neighbors \textit{o'} of the new node. The result of the range query is used to initialize the values of the new node's \textit{o.count\_after} and \textit{o.nn\_before}. If the size of \textit{o.nn\_before} is more than \textit{k} then the oldest timestamps are removed. For each \textit{o'}, the value of \textit{o'.count\_after} is increased by 1. Finally the new node is inserted into the ISB.
When a data point \textit{o} expires, meaning that \textit{o.t} is lower than the window's starting timestamp $W.start$, it is removed from the ISB. Its timestamp, however, is not removed from other nodes' list of preceding neighbors to mitigate overheads.

The above steps are applied in each slide. After they have been completed,  ISB is scanned for outliers. If a node's sum of \textit{count\_after} and the size of \textit{nn\_before}, whose timestamp is within the window borders, is lower than \textit{k}, then the node is an outlier. An optimization that avoids checking all objects in each slide is through the notion of \textit{safe inliers}: if a node's \textit{count\_after} is more than \textit{k}, then this node is a safe inlier and does not need to be checked again in future scans as it is guaranteed to have at least $k$ neighbors for the remainder of its lifetime.

In our parallel solution, we adopt both a structure for fast range queries (using, however, an M-tree instead of an ISB) and the notion of safe inliers.

\subsubsection{Thresh\_LEAP}
\label{sec:thresh_leap}
Thresh\_LEAP uses the broader notion of time-slicing \cite{WR09} in order to decrease the number of range queries needed to establish the status of each data point. Each window slide has a separate index structure where the data points are held. This design allows the use of the \textit{minimal probing principle}. Following this principle, each data point searches the succeeding slides' indices for neighbors and then continues with the preceding slides in reverse chronological order. This search stops when the data point has found \textit{k} neighbors.

Each data point \textit{o} has a list \textit{o.evil[]}, where it stores the number of neighbors in each preceding slide. Each slide has a \textit{trigger list}, where the data points with neighbors within the slide are stored.
A sketch of the main algorithm steps is as follows. For each new data point \textit{o}, a range query is issued on the same slide structure. If the search returns less than \textit{k} neighbors, range queries are issued for each preceding slide in reverse chronological order until \textit{k} neighbors are found or all slides are searched. The data point \textit{o} has its metadata updated with the number of neighbors for each slide and is inserted into the \textit{trigger lists} of all the slides where it has at least one neighbor. When a slide expires, all the data points in its \textit{trigger list} are re-processed. Each data point \textit{o} in the \textit{trigger list} issues a range query for the succeeding slides that have not been searched so far. There is no need to issue range queries on preceding slides as they have been searched in the previous steps of the algorithm.
After the previous steps have been completed, the status of each active data point \textit{o} is evaluated by summing up the content of the \textit{o.evil} list with its succeeding neighbors, specified in \textit{o.count\_after}.

In our solution, we use the time-slicing notion with the \textit{minimal probing principle} in order to decrease the number of range queries per data point.

\subsubsection{MCOD}
\label{sec:mcod}
The main motivation behind MCOD relates to the fact  that range queries are less expensive  than brute-force (all-pairs) distance computations but are, nonetheless, still expensive. MCOD (standing for Micro-cluster-based Continuous Outlier Detection) mitigates the need for range queries by drastically reducing the number of data points that need to be addressed during a range query through the creation of micro-clusters and the assignment of data points to them. A micro-cluster has at least \textit{k + 1} data points all of which are neighbors to each other. Its center can be a data point or just a point in the metric space and has a radius of $\frac{R}{2}$, implying that the maximum distance between any two objects in the micro-cluster is at most $R$. Each data point in any micro-cluster is an inlier and does not need to be checked in outlier detection queries. However, a data point that does not belong to a micro-cluster can be either an inlier or an outlier. Such objects are stored in a list $\mathbb{PO} \subseteq \mathbb{O} $.

On average, MCOD stores less metadata for each object than exact-Storm. More specifically, for each \textit{o} in a micro-cluster, it stores the identifier of its cluster. For each \textit{o} in $\mathbb{PO}$, it stores the \textit{o.count\_after} and the expiration time of the \textit{k} most recent preceding neighbors. MCOD also uses an event queue to store unsafe inliers that are not in any cluster. This event queue is a specific priority queue that keeps the time point at which a non-safe inlier should be re-checked.

A sketch of the algorithm steps is as follows. For each new data point \textit{o}, if \textit{o} is within $\frac{R}{2}$ range of a micro-cluster, it is added to it; if there are multiple such micro-clusters, the closest one is picked. Otherwise, if it has at least \textit{k} neighbors in $\mathbb{PO}$ within a distance of $\frac{R}{2}$, it becomes the center of a new micro-cluster. If none of the above conditions are met, \textit{o} is added to  $\mathbb{PO}$ and possibly to the event queue, if it is not an outlier. At each slide, all the previous non-expired outliers are checked along with the inliers for which the check time has arrived  (with the help of the event queue). When a data point \textit{o} expires, it is removed from the micro-cluster or $\mathbb{PO}$ and the event queue updates the unsafe inliers. If \textit{o} is removed from a micro-cluster and the points  remaining in that micro-cluster are less than \textit{k + 1}, then the cluster is destroyed and each data point of the cluster is processed as a new data point, without however updating their neighbors.

In our final parallel solution, we also adopt the notion of micro-clusters.

\section{Simple Solutions}
\label{sec:simple}

The aim of this work is to build upon the state-of-the-art non-parallel techniques and manage to parallelize the sliding window and the associated workload efficiently.
First, to explain the engineering approach chosen and to be able to assess the efficiency of the implementation of the parallel streaming solutions for distance-based outlier detection proposed in this paper, we introduce a baseline approach, which broadly corresponds to a single-partition implementation of exact-Storm in Flink. By single-partition, we mean that the  logical window is physically allocated to a single Flink node as a whole; by default, in Flink, the windows are physically partitioned across multiple Flink nodes. Then, we proceed to its parallelization, where we are forced to employ the notion of a \emph{meta-window}.

\subsection{A baseline approach in Flink}
\label{sec:baseline}

In this approach, we use two main components, as follows: (i) a \emph{stream handler}; and (ii) a \emph{window processor}. The stream handler applies a \texttt{map} function on each stream object and sends it to the window processor.
The window processor runs an outlier detection algorithm in each slide.

In the \texttt{map} function of the stream handler, each data point is initialized with a null \textit{o.count\_after} count and an empty list \textit{o.nn\_before}. These
records are sent to the single-partitioned window. The window has its own state in which it stores the records and is persistent across slides. In other words, changes made to a data point's metadata in a slide are kept throughout the data point's lifetime. Overall, the contents of a window at any point in time include all the active points along with their metadata, i.e., \textit{o.id}, \textit{o.value}, \textit{o.t}, \textit{o.count\_after} and \textit{o.nn\_before}.

The outlier detection algorithm contains two steps. In the first step, the update of each data point's metadata is performed. In the case of this algorithm, this is done only for the new arrivals. For every such data point, \textit{o}, it finds the neighbors, \textit{o'} in range \textit{R}, checking all the points in the window;  by default, the range is calculated based on the euclidean distance, but any type of metric distance can be employed.  If \textit{o'} is in the same slide as \textit{o}, then \textit{o.count\_after} is increased; otherwise the timestamp \textit{o'.t} is added to the list \textit{o.nn\_before}. For each \textit{o'} the value of \textit{o'.count\_after} is increased by 1. The remaining metadata associated with the older data points, i.e. the  \textit{o'.nn\_before} values,  have already been computed when these objects were inserted in the window, and so, do not need to be recalculated. The second step of the algorithm is to detect the outliers. For each data point in the current window, the algorithm computes the total number of neighbors. This is done by summing the \textit{o.count\_after} and the size of the list \textit{o.nn\_before}, taking into account only the values \textit{t} where $\textit{t} \geq \textit{W.start}$, as explained previously.

Both time-based and count-based windows are naturally supported in Flink. In this work, we mainly focus on time-based windows, but it is worth pointing out that, even without explicit support of count-based windows, it is straightforward to emulate them through artificially tweaking the initial timestamps, so that a fixed number of objects arrive and expire in each slide, and thus the number of alive objects remains constant during the stream processing.

\begin{figure}[tb!]
\begin{center}
\includegraphics[width=0.6\linewidth]{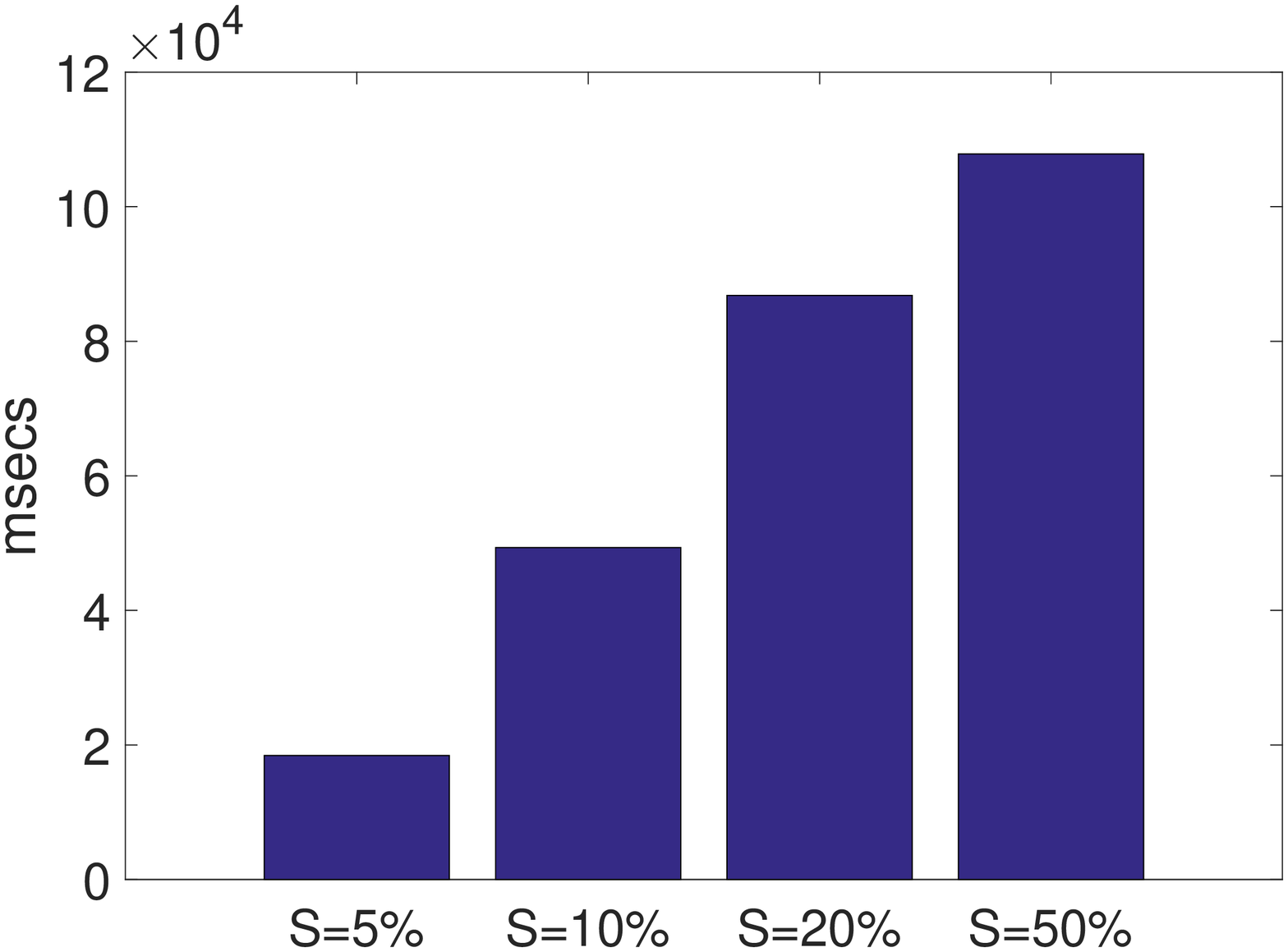}
\includegraphics[width=0.6\linewidth]{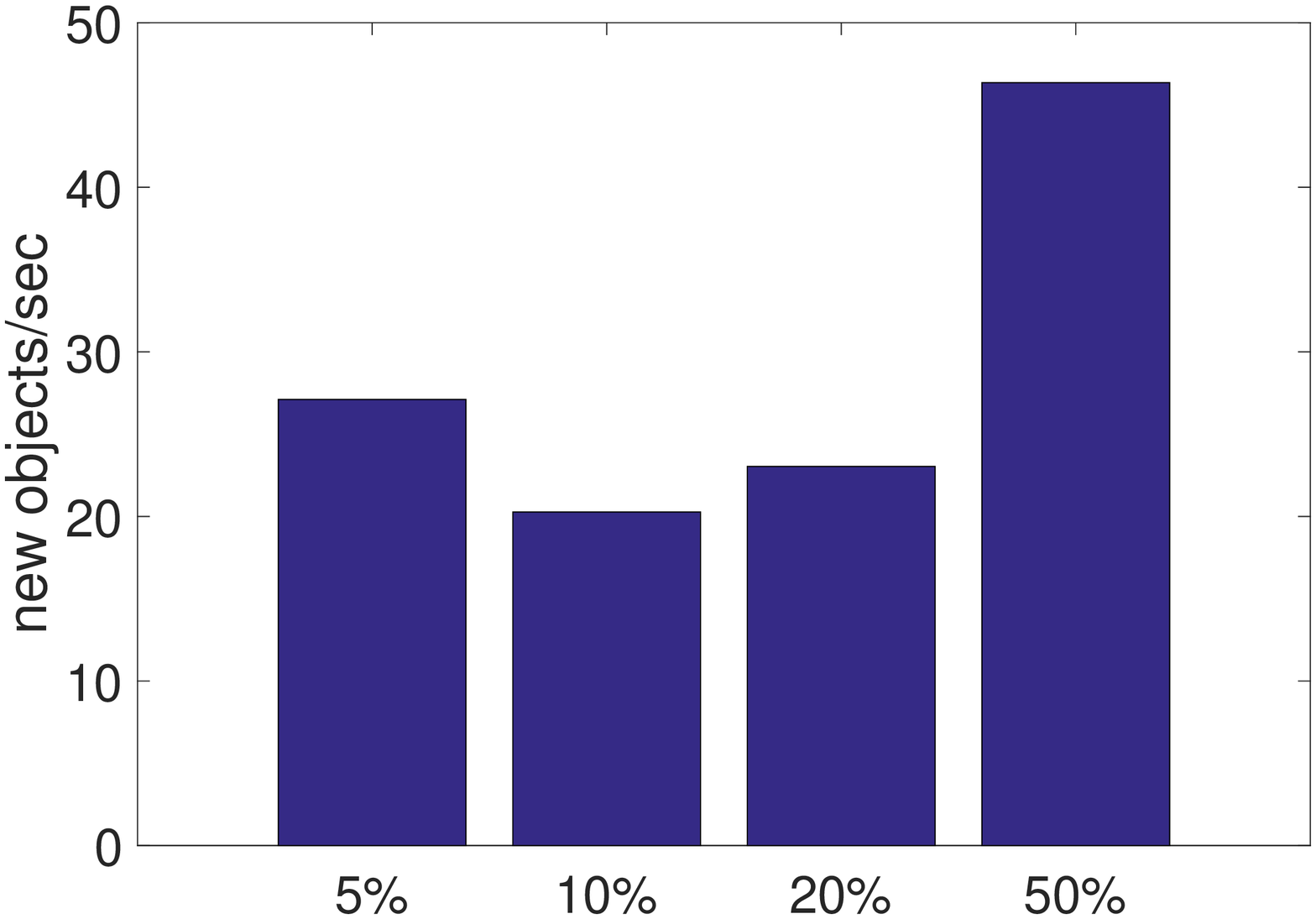}
\end{center}
\caption{Average processing time per slide (top) and input stream consumption rate (bottom) for the baseline approach.}
\label{fig:baseline}
\end{figure}

To gain insights into the performance of the solution, we evaluate this baseline approach using the Stock dataset from \cite{TranFS16}. It is a one-dimensional dataset with 1,048,575  data points\footnote{available from \url{https://wrds-web.wharton.upenn.edu/wrds}}. Each data point, \textit{o}, is assigned a unique identifier, \textit{o.id}, and has a numeric value of type \texttt{Double}, \textit{o.value}.  We employ a machine with an Intel i7-3770K CPU at 3.5GHz, which possesses 4 cores (8 threads) and 32GB of RAM. Figure \ref{fig:baseline} (top) shows the average processing time for each slide step considering four different values for slide magnitude, $S$, with $W=10K$; $S$  is stated as a percentage of $W$. Figure \ref{fig:baseline} (bottom) shows the corresponding input consumption rate, which is equal to the maximum streaming data throughput that  the baseline approach can support; in these settings, this throughput can reach 50 objects/sec. We can see that, in general, the average processing time per new arrival increases for both small slides, where a few new points arrive, and relatively large ones, where most of the window contents are replaced.


\begin{figure} [tb!]
	\centerline{\includegraphics[width=0.65\linewidth]{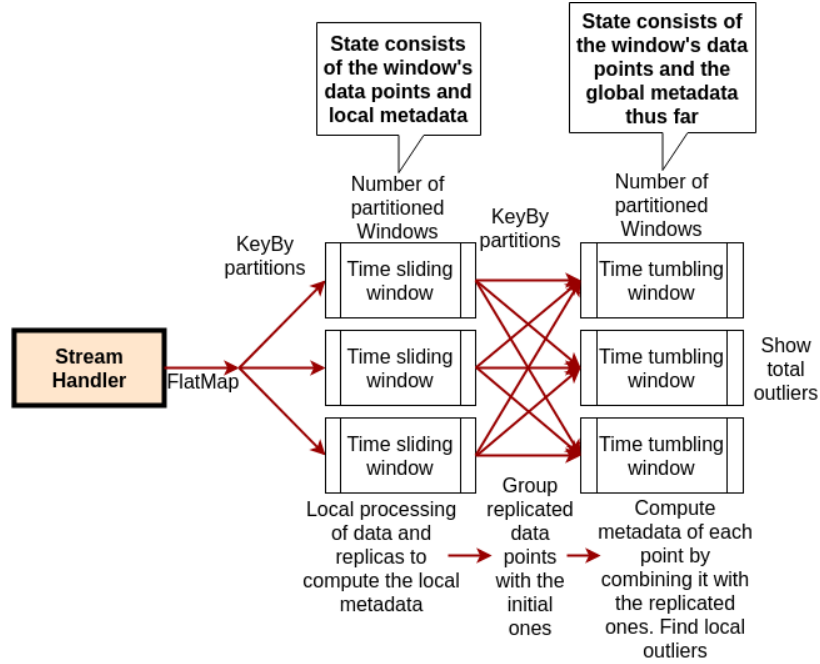}}
	\caption{Main implementation rationale with random partitioning, where the parallelized sliding window is followed by a partitioned tumbling meta-window.}
	\label{fig:basic}
\end{figure}

\subsection{A Naive Solution}
\label{sec:naive}

The parallelization of the baseline approach in Section \ref{sec:baseline} yields a naive parallel solution, where the window is split into a set of multiple partitions, $P$. It is deemed naive
in the sense that it does not benefit from any data structures to speed-up range queries. Notwithstanding its simplicity, the parallelization technique needs to efficiently address the challenges relating to (i) the collaboration between physical window partitions to establish whether a point, at a specific time, is an outlier or not by aggregating local statistics; and (ii) the maintenance of the window state across slides, where the state includes the object metadata.

The devised engineering solutions  need to respect the principle that, in each window slide, each new object is processed only once. This does not allow first to compute the local aggregates for a given point and then to compute the global aggregates into the same window. Therefore,
the key idea is to split the window processor into two parts, the sliding window processor and the tumbling window processor, as shown in Figure \ref{fig:basic}. The former holds the active points in its partition, allowing some temporary replication, as discussed in the following, while the latter keeps the final metadata in each slide. These metadata also include information about detected outliers. Since the metadata evolve in each slide for both  new and old points, the window state gets fully updated, and thus the tumbling window semantics apply. Essentially, the second window serves as a \emph{meta-window}.

\begin{algorithm} [tb!]
\caption{Naive solution}\label{alg:naive}
\begin{algorithmic}
\small
\Procedure{StreamHandler}{}
\For{each~new~object $o$}
\State \text{initialize record}
\If{there~is~no~timestamp}
\State \text{add~artificial~timestamp}
\EndIf
\State $o.partition \gets o.id$ \texttt{mod} $|P|$
\State $o.flag \gets 1$
\For{each partition $p \in P$}
\If{($p==o.partition$)}
\State $o.flag \gets 0$
\State  \text{send~$o$~to~$p$}
\State $o.flag \gets 1$
\Else
\State  \text{send~$o$~to~$p$}
\EndIf
\EndFor
\EndFor
\EndProcedure
\State
\Procedure{SlidingWindowPartitionProcessor}{}
\For{each~slide}
\State \text{evict expired objects and old objects with $o.flag$==1 }
\State \text{compute pairwise distances involving new objects}
\State \text{update $o.count\_after$ and $o.nn\_before$ metadata}
\For{each~object $o \in \mathbb{PO}$}
\If{($o.count\_after \ge k$) )}
\State  $\mathbb{PO} \gets \mathbb{PO} \setminus o$
\EndIf
\EndFor
\State \text{group objects in  $\mathbb{PO}$ by $o.partition$ }
\State \text{and send to a (new) tumbling window}
\EndFor
\EndProcedure
\State
\Procedure{TumblingWindowPartitionProcessor}{}
\For{each~slide}
\State \text{aggregate $o.count\_after$ and $o.nn\_before$ metadata}
\For{each~object $o$}
\State $o.nn\_ prec\gets$ \text{prune} $o.nn\_before$ 
\If{($o.count\_after+  |o.nn\_prec| < k$) )}
\State  \text{report~$o$~as~an~outlier}
\EndIf
\EndFor
\EndFor
\EndProcedure
\end{algorithmic}
\end{algorithm}

The implementation details are as follows.
First, we extend the object record with two new fields, namely $o.flag$ and $o.partition$. The former is a binary variable, where 0 indicates that the object should be kept to the assigned window during its lifetime, and 1 indicates that the object is redundant and, therefore, should be evicted in the next slide, regardless of the $W.start$ value. The stream handler uses a \texttt{flatMap} for not only initializing an object's extended record, but also to compute the object's  $o.partition$ based on the $o.id$. Then, it dispatches the object to \emph{all} partitions, setting the $o.flag$ to 1 to all partitions different to $o.partition$. In other words,  new objects are replicated across all the partitions.

The sliding window processor is responsible for comparing the distances between a) the new objects in the slide and b) the new objects and all the previous window contents, the timestamp of which comes after $W.start$. This leads to the need for updating the $o.count\_after$ value for all objects and the $o.nn\_before$ value for the new objects only. However, because the metadata for the new objects are local aggregates spread across all partitions, to produce the global aggregates, the updated objects are partitioned again according to $o.partition$ into a new tumbling window, but without replication this time.   To save communication cost, not all objects are shuffled, but only those that are not safe inliers. The set of the safe inliers is the complement of the set of potential outliers, i.e., $\mathbb{O}\setminus \mathbb{PO}$.
In each slide, the sliding window processor first creates the $\mathbb{PO}$  set by checking whether $o.count\_after$ is less than $k$ or not.
Then, the tumbling window globally aggregates the local aggregates of non-safe inliers thus producing the exact metadata needed to establish outlierness. As in  \cite{AF07}, the list in $o.nn\_before$ may  contain preceding neighbors that have expired, and so, a filter is required to identify the alive ones, termed $o.nn\_prec$.

Algorithm \ref{alg:naive} summarizes the naive approach. In conclusion, the naive solution requires minimal effort from the stream handler at the expense of high communication and computation cost, since new objects are sent to each partition and compared against the full local window contents. In terms of implementation, it requires the notion of \emph{meta-windows} to achieve accurate results.


\section{Advanced Solution}
\label{sec:advanced}

The advanced solution extends the naive one in two complementary and orthogonal dimensions, namely thr-ough: (i) employing data structures to support fast range queries and processing of window elements, described in Section \ref{sec:wsm}; and (ii) through performing value-based partitioning, which eliminates the need to employ a meta-window, described in Section \ref{sec:vbp}.

\subsection{Window state management}
\label{sec:wsm}

We can enhance the processing of new and old items in the partitioned window using two techniques.
The first technique involves a more advanced approach to holding state in the sliding window, by storing it in a M-tree \cite{CPZ97}, against which range queries are submitted.
M-trees are a part of each partition state, therefore each Flink partition has its own local tree allowing  faster execution of local range queries.

An additional enhancement derived from ideas described in \cite{cao2014scalable} and introduces the notion of time-slicing in the window's state, where a  window is divided into a number of slides and each slide has its own index structure, replacing the one structure associated with the whole window. This design allows the application of the \textit{minimal probing principle}, as discussed in Section \ref{sec:thresh_leap}, having as an index structure in this case the M-tree which allows, as described above, faster range queries. This means that a new data point  first issues  a range query on the its own slide's M-tree, and, if the search returns less than \textit{k} neighbors, it continues issuing range queries on the previous slides' M-trees. The list of the preceding neighbors on the metadata of the data points is replaced by a list containing the number of neighbors for each preceding slide. Finally, each slide has a \textit{trigger list} containing all of the data points that will be affected when the slide is expired.

Note that, compared to Algorithm \ref{alg:naive}, the key difference is in the sliding window processor.

\subsection{Value-based partitioning}
\label{sec:vbp}

\begin{figure} [tb!]
	\centerline{\includegraphics[width=0.55\linewidth]{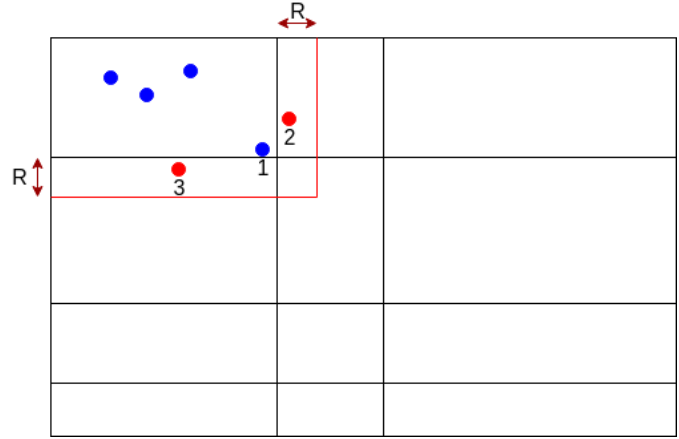}}
	\caption{Value-based partitioning}
	\label{fig:value-based-part}
\end{figure}

A limitation of the solutions described thus far is that they replicate each new data point to all partitions. This is inevitable, given that the stream handler assigns points to partitions randomly and thus a new object may have neighbors in all partitions. Value-based partitioning addresses this limitation without sacrificing the accuracy of the results. Also, as will be explained below, it eliminates the need to exchange information between partitions during a slide.

\subsubsection{Grid-based partitioning}
As a starting point, we make an additional assumption that the space is  Euclidean and, therefore, it can be partitioned into grid cells. Further, we assume  the availability of sample data before execution. Based on the sample data, we can extract min, max and quantile information about the value distribution in each dimension can be extracted, in order to construct the grid cell appropriately.

The rationale behind the approach for a two-dimensional case is illustrated in Figure \ref{fig:value-based-part}.
Each cell is assigned to a single Flink node. However, an object in a cell may have neighbors in other cells as well. The key difference is that these neighbors belong to adjacent cells only; therefore the number of Flink nodes that need to be aware of the arrival of each new object is limited. More specifically, the borders of each cell are extended by a buffer zone of width equal to $R$. The stream handler sends a new data point (i) to the partition corresponding to its cell with flag $o.flag$ set to 0 and (ii) to all the partitions, the buffer zone of which includes the new data point with flag $o.flag$ set to 1; these partitions form the set $AP$ in Algorithm \ref{alg:advanced-vbp}.  Assuming that $R$ is much smaller than the size of a grid cell side, each data object is replicated at most 4 times if the data is 2-dimensional;  this is because, when it falls near to a cell corner, it may fall into the buffer zone of three other adjacent cells. In the example in Figure \ref{fig:value-based-part}, the buffer zone borders for the upper-left cell are depicted; points 1,2 and 3 are sent to the Flink node responsible for the upper-left cell along with all the other three points. For $d$-dimensional data, a data point can be sent to up to $2^d$ partitions, which grows exponentially in the number of dimensions but, by construction, is at most equal to $|P|$; i.e., the replication is never inferior to that of the naive solution.

\begin{algorithm} [tb!]
\caption{Advanced solution with value-based partitioning}\label{alg:advanced-vbp}
\begin{algorithmic}
\small
\Procedure{StreamHandler}{}
\For{each~new~object $o$}
\State \text{initialize record}
\If{there~is~no~timestamp}
\State \text{add~artificial~timestamp}
\EndIf
\State $o.partition \gets $ \texttt{findGridCell} $(o.value)$
\State $o.flag \gets 0$
\State  \text{send~$o$~to~$o.partition$}
\State $AO \gets $ \texttt{findRelevantAdjacentPartitions} $(o.value)$
\State $o.flag \gets 1$
\For{each partition $p \in AP$}
\State  \text{send~$o$~to~$p$}
\EndFor
\EndFor
\EndProcedure
\State
\Procedure{SlidingWindowPartitionProcessor}{}
\For{each~slide}
\State \text{evict expired objects from M-tree }
\State \text{insert new objects in M-tree }
\State \text{compute distances}
\State \text{update $o.count\_after$ and $o.nn\_before$ metadata}
\For{each~object $o \in \mathbb{PO}$}
\If{($o.count\_after \ge k$) )}
\State  $\mathbb{PO} \gets \mathbb{PO} \setminus o$
\Else
\State $o.nn\_ prec\gets$ \text{prune} $o.nn\_before$ 
\If{($o.count\_after+  |o.nn\_prec| < k$) )}
\State  \text{report~$o$~as~an~outlier}
\EndIf
\EndIf
\EndFor
\EndFor
\EndProcedure
\end{algorithmic}
\end{algorithm}

According to the partitioning above, each partition has all the necessary information in order to establish object outlierness locally. Therefore, a single sliding window partition processor is required, which incorporates the responsibility of the tumbling window partition processor in the previous approaches. Algorithm \ref{alg:advanced-vbp} summarizes the advanced solution with value-based partitioning (referred to as \emph{advanced(grid)}  in the experiments). Apart from making the tumbling window processing phase obsolete, another difference to the simple advanced algorithm is that objects with $o.flag =1$ are not dropped until they expire.

In a streaming case, data is inherently volatile. However,
dynamic load balancing is left for future work. As shown in Figure \ref{fig:value-based-part}, the grid cells are not necessarily of the same size. Here, we define cell boundaries statically, taking into consideration only the value distribution per dimension, thus overlooking issues such as actual computation and communication cost per partition.

\subsubsection{Tree-based partitioning}
The main drawback of grid-based partitioning is that it does not support arbitrary metric spaces. To address this limitation,
we can use the M-tree also to perform value-based partitioning.
In particular, instead of assuming a grid partitioning, which presupposes an Euclidean space, we can partition the data based on a rather shallow level of the M-tree, as follows. First, fix a particular level $\ell$ of the M-tree. Then, each Flink node is assigned to a particular node (or multiple nodes) at level $\ell$. Level $\ell$ depends on the choice of $R$. However, the overlap between nodes at level $\ell$ depends on the distribution of the data objects.
Also, the tree needs to be initially populated with enough elements of the data stream so that it grow at least up the level $\ell$.

The main guideline in implementing a structure such as the M-tree for the partitioning is that the replication of the data points should not significantly exceed the replication from the grid partitioning.
The M-tree splits the nodes in such a way that the area covered by each node overlaps with the other nodes. The extent of the overlaps is
independent of  \textit{R} and is based only on the quality of the promotion and the partitioning functions used in the tree constructions. In our initial experiments, the partitioning based on a M-tree yielded a replication rate for each data point close to 1.5 to 2 times more than the corresponding rate in the grid partitioning.

Another type of tree that provides fast range queries in a metric space is the Vantage-Point tree (VP-tree) \cite{yianilos1993data}. VP-tree works by selecting a vantage point from the dataset and a corresponding threshold for each node. After the selection, the data points are split into two children nodes, the ones near the vantage point and in radius of the threshold, and the ones outside the threshold. This means that the tree is binary and there is no overlapping involved.
Just as the partitioning with the M-tree, in the VP tree solution, each Flink node is assigned to a particular (or multiple) tree nodes at level $\ell$. Because the tree is binary, the level can be selected based on the number of Flink partitions, e.g. the level for 16 Flink partitions should be at least 4 (not counting the root level). In our initial experiments, this solution exhibited much less replication than the M-tree one, with a replication rate closer to that of the grid partitioning. This solution is referred to as \emph{advanced(tree)} in the experiments. Similarly to M-trees, VP-trees are sensitive to their construction parameters, too.

Overall, the advanced solution trades i) additional workload on the stream handler and ii) increased memory requirements on the Flink nodes for i) less communication cost between both the stream handler and the Flink nodes, and the nodes themselves, and ii) less computation cost per node. The implementation challenges are mostly related to how the value-based partitioner on the stream handler is efficiently and effectively constructed, while, in the experiments, we assess the impact of the structures for state management.

\section{pMCOD}
\label{sec:pmcod}

As described in Section \ref{sec:mcod}, the \emph{MCOD} algorithm is the best non-parallel solution for the distance-based outlier detection problem according to the results in \cite{TranFS16}. The \emph{MCOD} algorithm has three distinct features that could be independently implemented in the final parallel solution termed as \emph{pMCOD}:
\begin{itemize}
	\item 	the micro-clusters;
	\item	the event-queue; and
	\item	restricted usage of the M-tree for the range queries.
\end{itemize}

Each of these features is  examined in-depth for its usefulness to the final solution hereby, and the accompanying experiments are in Section \ref{sec:pmcod-features}.

\begin{algorithm} [tb!]
\caption{pMCOD}\label{alg:pMCOD}
\begin{algorithmic}
\small
\Procedure{StreamHandler}{}
\State \text{Same as Algorithm 2}
\EndProcedure
\State
\Procedure{SlidingWindowPartitionProcessor}{}
\For{each~slide}
\State \text{evict expired objects from M-tree }
\If{a micro-cluster dissolves}
\State \text{treat all points as new w/o updating their neighbors}
\EndIf
\State \text{insert new objects in M-tree }
\For{each new object $o'$ with $o'.flag=0$ }
\If{$o'$ belongs to micro-cluster}
\For{each object $o'' \in \mathbb{PO}$}
\State \text{update $o''$ metadata due to $o'$}
\EndFor
\Else
\State $\mathbb{PO} \gets \mathbb{PO} \cup o'$
\State \text{compute pairwise distances involving $o'$}
\State \text{check if a new micro-cluster can be formed }
\EndIf
\EndFor
\For{each~object $o \in \mathbb{PO}$}
\If{($o.count\_after \ge k$) )}
\State  $\mathbb{PO} \gets \mathbb{PO} \setminus o$
\Else
\State $o.nn\_ prec\gets$ \text{prune} $o.nn\_before$ 
\If{($o.count\_after+  |o.nn\_prec| < k$) )}
\State  \text{report~$o$~as~an~outlier}
\EndIf
\EndIf
\EndFor

\EndFor
\EndProcedure
\end{algorithmic}
\end{algorithm}

\subsection{Micro-clusters}
\label{sec:pmcod-mc}

The first version of the \emph{pMCOD} algorithm combines the value-based partitioning and the M-tree with the micro-clusters as shown in Algorithm \ref{alg:pMCOD}. The motivation behind using micro-clusters is to drastically reduce the number of range queries submitted to the M-tree, as explained in Section \ref{sec:mcod}. In contrast to the work in \cite{KontakiGPTM16}, this version does not contain an event queue. The sliding window's state consists of the micro-clusters, the potential outliers $\mathbb{PO}$ and the M-tree. The notion of value-partitioning from Section \ref{sec:advanced} along with the introduction of micro-clusters means that each partition is able to fully report its outliers at a faster rate without the need to communicate with the other partitions.

Each window slide starts with the eviction of the expired data points from its state and the dissolution of the micro-clusters with $\le k$ elements. Each data point that belonged to a dissolved micro-cluster is treated as a new data point. For each new data point, the algorithm computes its distance to the micro-clusters and if it belongs to any of them, it proceeds to updating the $\mathbb{PO}$ metadata only. If the data point does not belong to any micro-cluster, it is inserted into the $\mathbb{PO}$ set and a range query is executed to find its neighbors. Based on the number of neighbors of a data point in $\mathbb{PO}$, a new micro-cluster may be created.

Broadly, the set of potential outliers includes only points that do not belong to a micro-cluster. Also, if a point belongs to a micro-cluster, only the metadata of points in $\mathbb{PO}$ need to be updated.
After the update of each data point's metadata, the algorithm reports the outliers by checking the data points in $\mathbb{PO}$. The list $o.nn\_before$ is pruned to contain only the non-expired objects. Each data point for which $o.count\_after ~+$ ~$|o.nn\_before| < k$ holds, is reported as an outlier for the corresponding slide.

\subsection{Event queue}
\label{subsec:eventq}

The motivation behind the event queue is to quickly re-process data points that are affected by the expiring slide without unnecessary checks of the points in $\mathbb{PO}$; it forms the cornerstone of the \emph{COD} algorithm in \cite{KGP+11,KontakiGPTM16}. Every data point with a preceding neighbor is inserted in the event queue along with the timestamp of the oldest neighbor. The queue is essentially a priority queue, in the sense that the data point that has the oldest neighbor in the active window will be the first data point in the structure that makes sense to check.

Such an event queue is employed in two places. First, during the insertion of a data point, when a data point is being processed by the current window, the timestamp of its oldest neighbor is retrieved and the data point along with that timestamp are saved in the queue. Second, when a slide expires, the first item of the queue is checked to identify as to whether the timestamp belongs to or is older than the expiring slide; if this is the case, the data point needs to be re-processed in order to define its status. If the timestamp is more recent than the expiring slide, then the process stops.

The implementation of the event queue in the \emph{pMCOD} algorithm has been done in two separate ways. The first implementation uses the \textit{Priority Queue} data structure, while the second uses the \textit{TreeSet} data structure in Java. In both implementations, the event queue is part of the window's state.

\subsection{Usage of M-tree}
\label{subsec:mtree}

In the \emph{MCOD} and \emph{pMCOD} algorithms, where the need for range queries has decreased, the use of the M-tree for faster range queries is under question, as to whether its overhead outweighs the benefits it incurs. To investigate this issue, we have implemented and evaluated four different implementation schemes.

In the first flavor, as described in Section \ref{sec:pmcod-mc}, the M-tree is used to include all the data points of the window in each partition. This means that the tree structure includes both the data points in $\mathbb{PO}$ and the data points in the micro-clusters.

The second implementation does not use the M-tree at all. All  range queries are done using a custom distance function, such as the euclidean, on the data points of $\mathbb{PO}$ or on the centers of the micro-clusters.

The third implementation uses one M-tree that the data points in $\mathbb{PO}$. This means that, whenever a range query on the $\mathbb{PO}$ is issued, it is completed by the M-tree, while a range query on the micro-clusters centers is evaluated with the help of a custom distance function.

The forth and final flavor uses two different M-trees: one where the data points in $\mathbb{PO}$ are stored and one where the centers of the micro-clusters are stored. According to this implementation, every range query issued is processed through one of the two M-trees.

\section{Performance Evaluation}
\label{sec:eval}

The experimental setting is as follows. We focus on presenting the performance as a function of the window size, the slide size, the amount of outliers and the degree of parallelism. The accuracy is always 100\%, since all techniques are exact. We have employed three real and one artificial dataset.  These datasets are static and finite, but are adequate to emulate a streaming setting. More specifically, the techniques are implemented so that they can handle infinite streams. But in order to present repeatable results, we employ publicly available datasets to produce finite streams and, unless stated otherwise, the times presented correspond to the average time per slide, aggregated over 200 slides overall. Note that, given that we are in a streaming environment and we report times per slide, the actual full dataset size does not matter.

In addition, we do not activate time-slicing in the advance solutions, pMCOD flavor is as described in Section \ref{sec:pmcod-mc}, and the grid-based technique is used for value-based partitioning.  First, we present results using a single multi-core machine, the same machine as the one described in Section \ref{sec:baseline}.  Then we move to experiments using a small cluster. Initially, we focus on the Stock real-world dataset. Each experiment is repeated 5 times. $R$  and $k$ are set to 0.45 and 50, respectively, yielding 1.02\% of outliers (i.e., the setting is similar to \cite{TranFS16}). The default degree of parallelism, i.e., the number of Flink partitions of the window is 16, and each Flink node runs on a single core.
Stock is an one-dimensional dataset. 
To allow a fair comparison, the timestamps are assigned in such a way that all windows are of the same size, and the slide is given as a percentage of $W$, e.g., a slide of 5\% means that the 5\% of the window contents are new arrivals. 

\subsection{Algorithm Comparison and Evaluation on a single machine}

\begin{figure}[tb!]

\begin{tabular}{c}
	\centerline{\includegraphics[width=0.65\linewidth]{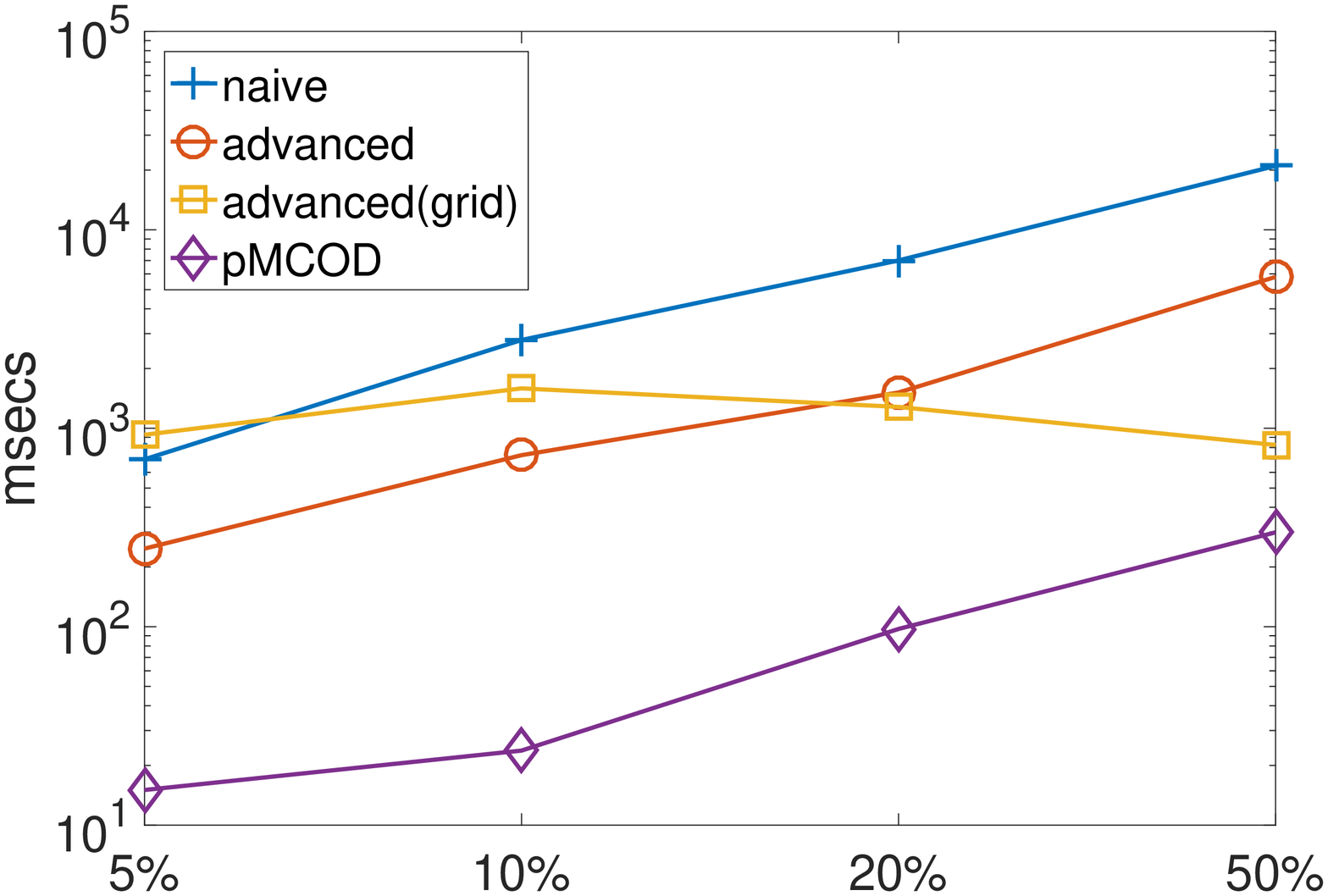}}~\\
	\centerline{\includegraphics[width=0.65\linewidth]{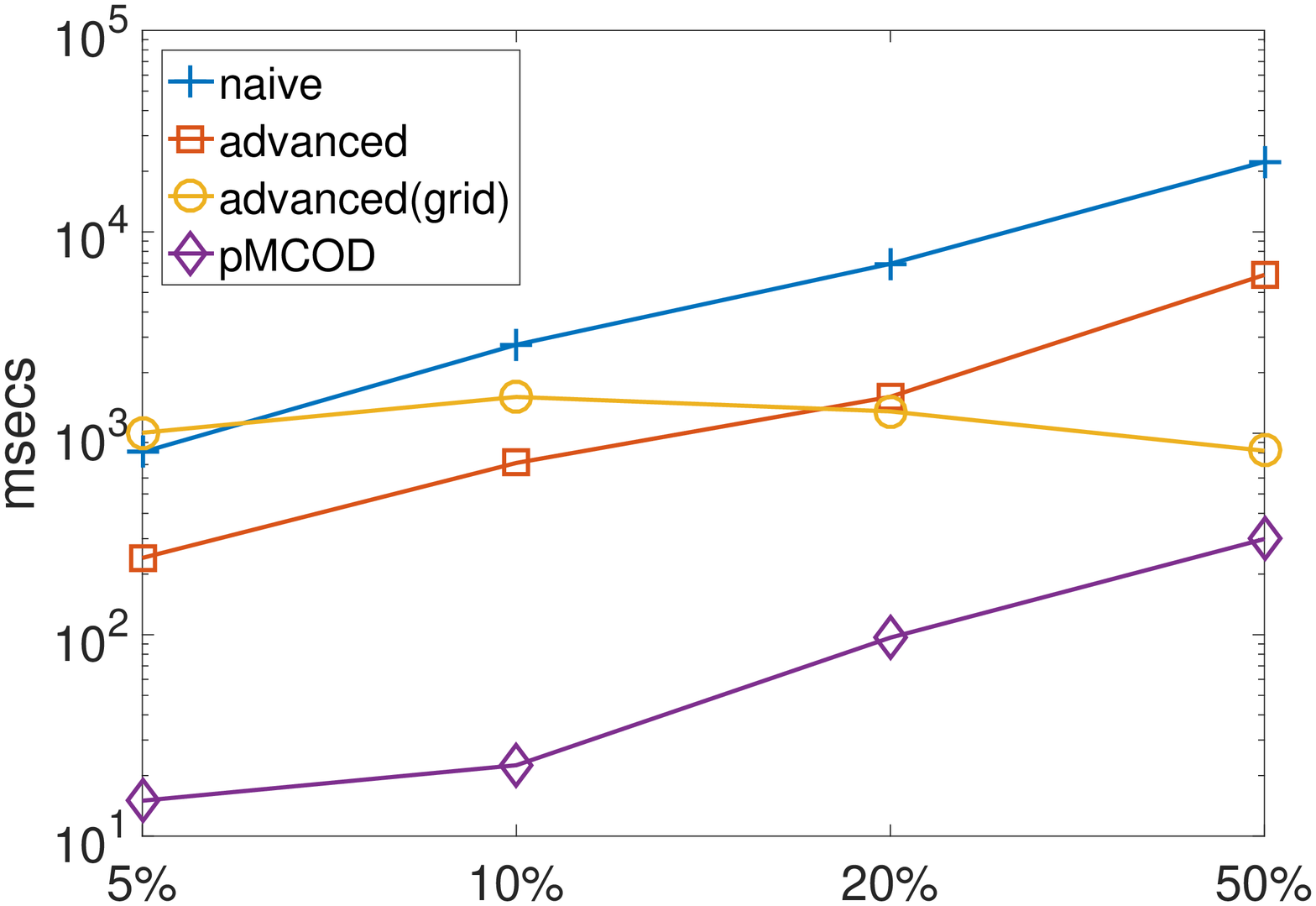}}
\end{tabular}
	\caption{Average (top) and median (bottom) processing time per slide  for different slides and window of 10K objects.}
	\label{fig:perf10K}
\end{figure}

\begin{figure}[tb!]
	\centerline{\includegraphics[width=0.65\linewidth]{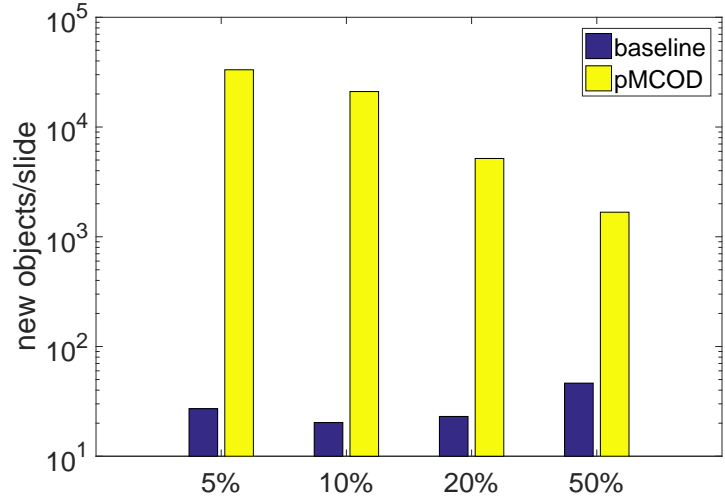}}
	\caption{Throughput comparison for $W=10K$ and slides of 5\%-50\%.}
	\label{fig:throughput}
\end{figure}

\begin{figure*}[tb!]
\begin{tabular}{ccc}
	\includegraphics[width=0.32\linewidth]{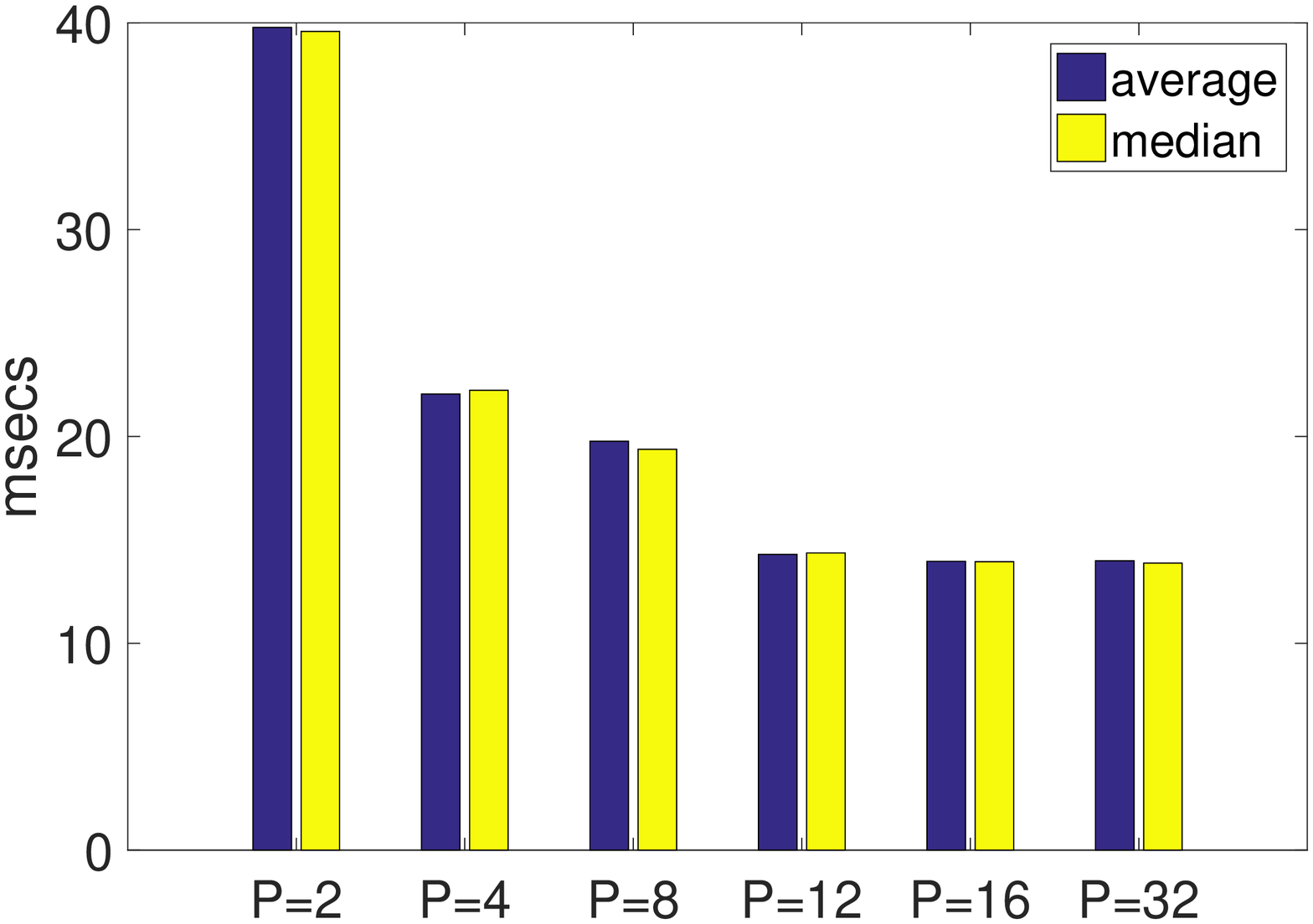} & \includegraphics[width=0.32\linewidth]{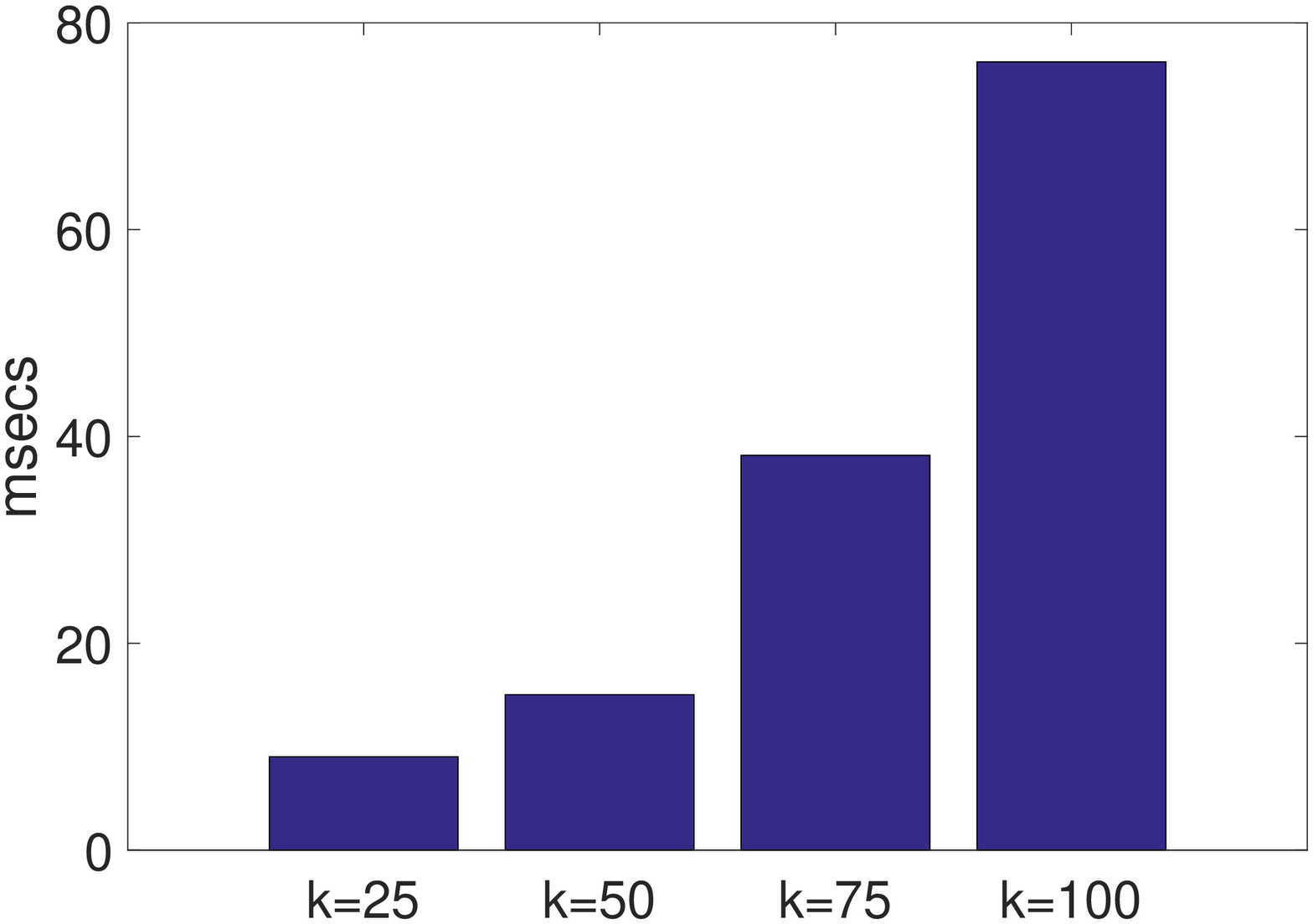} & \includegraphics[width=0.32\linewidth]{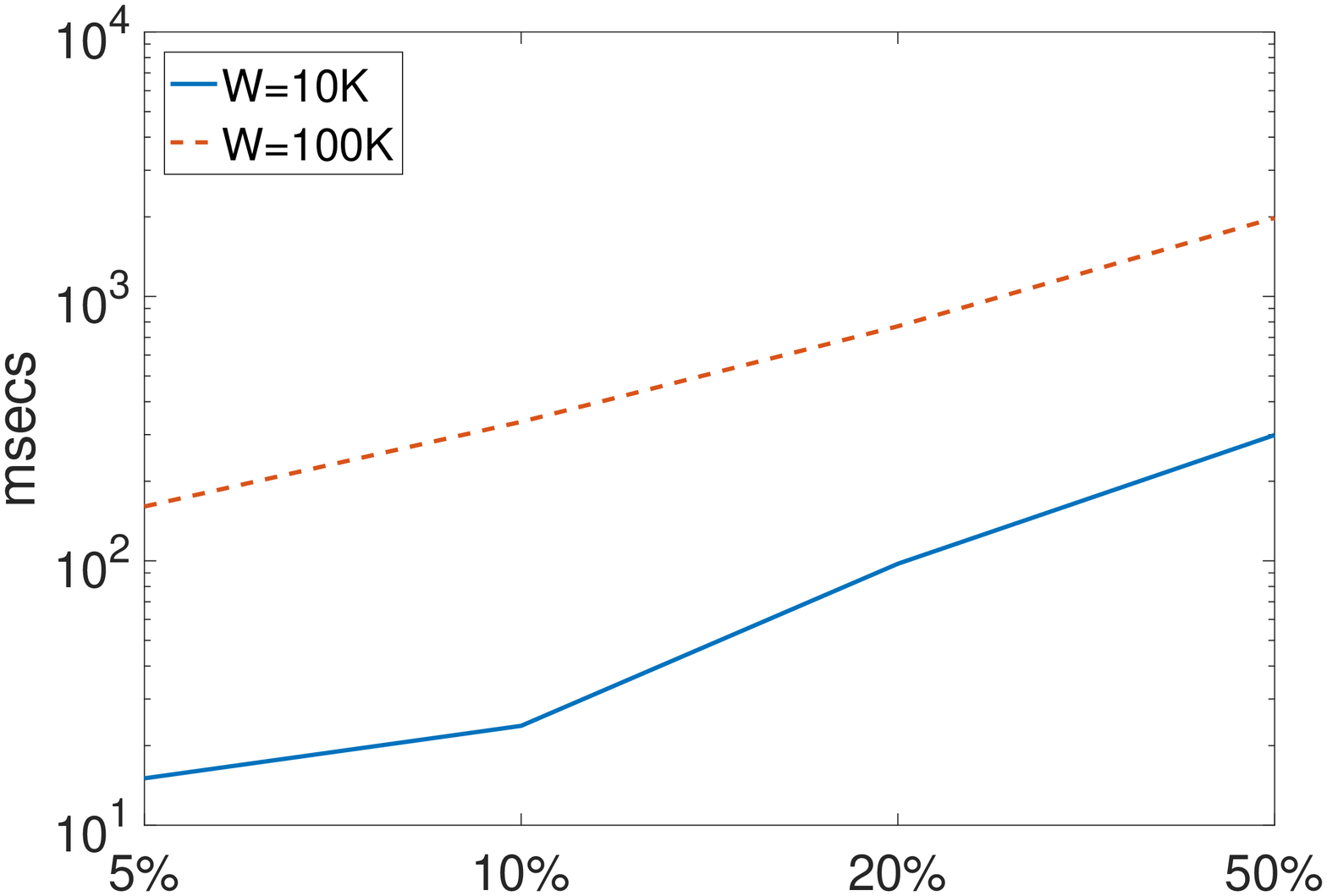} \\
\end{tabular}
	\caption{The impact of the degree of parallelism $P$ (left), $k$ (middle) and the window size $W$ (right - for slide sizes of 5\%-50\%) on pMCOD's performance.}
	\label{fig:pMCOD}
\end{figure*}

In the first experiment, we employ a window of 10K objects, while the slide varies from 5\% to 50\%.  The results are shown in Figure \ref{fig:perf10K}, where both the average and the median times are reported. From the figure, we can draw the following observations:
\begin{enumerate}
\item \emph{pMCOD} improves upon the \emph{naive} solution by
two orders of magnitude; for example, it is 117X faster for slide 10\%.
\item \emph{pMCOD} improves upon the advanced solutions with both  random and value-based partitioning (labeled as \emph{advanced} and \emph{advanced(grid)}, respectively) by an order of magnitude for slides up to 20\%; e.g., for slide of 10\%, it is 10X faster, and for 20\%, it is 13X.

\item For slides of 50\%, where half of the window points are new in each slide, \emph{pMCOD} is faster than \emph{advanced(grid)} by 2.74 times.

\item \emph{Advanced} dominates \emph{advanced(grid)} for small slides, while the latter is better for large ones, which is mostly attributed to the fact that \emph{advanced(grid)} inherent load imbalance\footnote{The grid cell used for partitioning is based on an initial sample and thus no guarantees can be provided as to how balanced the workload distribution is throughout stream processing.} is outweighed by the benefits of less replication and communication in large slides.
\item Despite the non-negligible standard deviation, the trends in the average values are similar to those in the median ones.
\end{enumerate}

Figure \ref{fig:throughput} compares \emph{pMCOD} against the  throughput results obtained by the baseline technique, described in Section \ref{sec:baseline}. The improvements are up to 2076 times, whereas \emph{pMCOD}'s throughput exceeds 33240 new objects per second for $S=5\%$. However, we observe that the throughput degrades as the slide increases.

In the second experiment, we focus on \emph{pMCOD} and we examine the impact of three parameters, namely the degree of parallelism $P$, $k$ and the window size $W$. The results are summarized in Figure \ref{fig:pMCOD}. Regarding the degree of parallelism, the left figure refers to a setting where $W$ is 10K and the slide is 5\%.
We see that \emph{pMCOD} scales well and the time per slide drops nearly two times when we go from two partitions to four.  Since the machine is a 4-core/8-thread one, the gains are small for higher degrees of parallelism. We also see that our default configuration of $P=16$ yields the highest performance. In the middle figure, we see that, as we increase the $k$ value, the performance degrades. Increasing the $k$ value implies more outliers and higher difficulty in forming micro-clusters; as such, this result is reasonable. Finally, the rightmost figure reveals that \emph{pMCOD} scales well with the size of the window: a ten-fold increase in the window size results in similar increases in the processing time for slides of 5\% and 10\%, and smaller increases for larger slides. More specifically, for 20\% slide magnitude, the increase in the processing time in the 100K window is less than 8 times, and for 50\% slide, it is 6.62 times only.

\begin{figure}[tb!]
	\centerline{\includegraphics[width=0.65\linewidth]{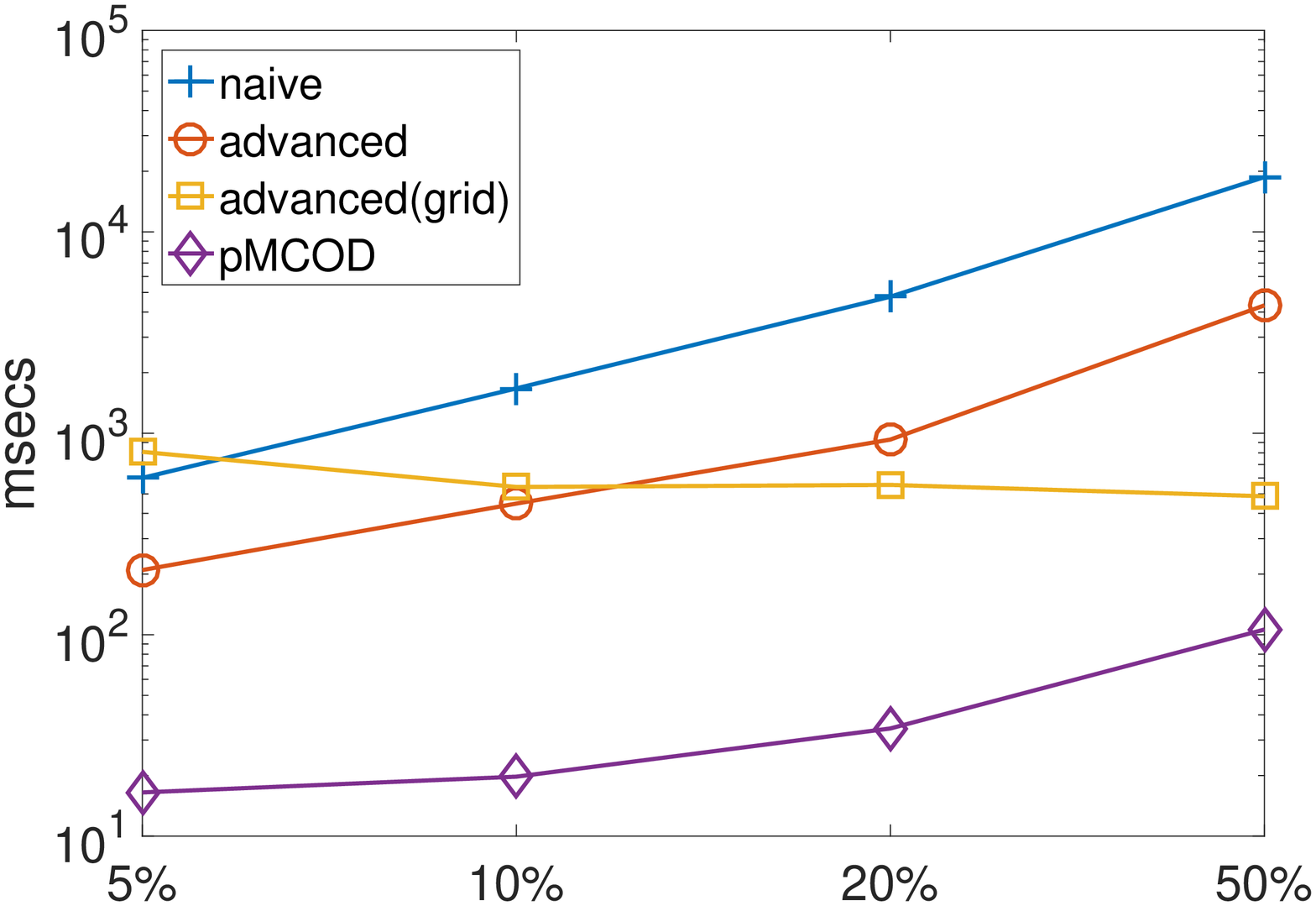}}
	\caption{Average  slide processing time for slide sizes of 5\%-50\% for the artifical dataset}
	\label{fig:gauss}
\end{figure}

\subsubsection{Using Additional Datasets with More Dimensions} First, we provide results using an artifical dataset, which is generated from a mixture of  three Gaussian distributions and taken from \cite{TranFS16}. We set $W=10K$, $R=0.28$  and $k=50$. Figure \ref{fig:gauss} presents the results, where it is shown that the main observations drawn for the Stock real-world dataset hold.

\begin{figure}[tb!]
	\centerline{\includegraphics[width=0.65\linewidth]{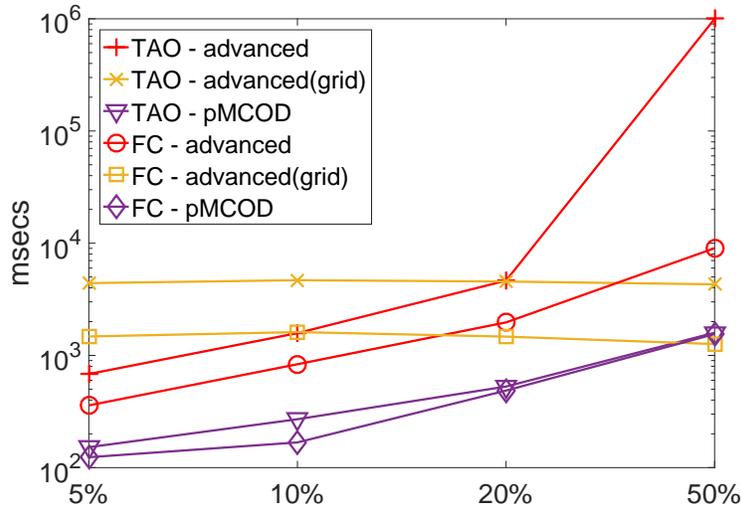}}
	\caption{Average  slide processing time for slide sizes of 5\%-50\% for the TAO and FC datasets}
	\label{fig:tao-fc}
\end{figure}

In the next experiment, we employ two additional real-world datasets, namely Forest Cover (FC)\footnote{Available from \url{http://kdd.ics.uci.edu}} and TAO\footnote{Available from \url{http://www.pmel.noaa.gov}}, considering 2 and 3 dimensions, respectively.
We configure their parameters so that always $k$ is 50, while $W$ is kept to 10K. More specifically, for FC, we set $R=34$ on the 2nd and the 5th dimensions (corresponding to 1.3\% outliers - we discard the other 53 dimensions), and for TAO, we set $R=1.9$ on all three dimensions (corresponding to  0.98\% outliers).

The results are shown in Figure \ref{fig:tao-fc} (actually, for TAO and S=50\%, \emph{advanced} crashed due to memory shortage).  The key observations are as follows: (i) \emph{pMCOD} behavior is nearly the same for 2 or 3 dimensions but significantly worse than the behavior for the one-dimensional datasets; the latter is attributed to the known performance degradation of M-tree for higher dimensions and the increased replication; (ii) for the FC dataset, and due to the increased number of outliers compared to the other settings, \emph{advanced(grid)} slightly outperforms \emph{pMCOD} when $S=50\%$, which means that half of the points in each window slide are new arrivals; and (iii) \emph{advanced} and \emph{advanced(grid)} are significantly affected by the increase in the number of dimensions considered from 2 to 3.

Working with higher number of dimensions than 3 is left for future work, since distance-based outlier detection is known to be problematic in such cases due to the curse of dimensionality issues \cite{Aggarwal2013}. The solutions for high-dimensional settings employ a different approach, such as other metrics for outlier definition, approximations, dimensionality reduction and subspace clustering, which are out of our scope.

\subsubsection{Comparison against results in \cite{TranFS16}}

\begin{table}[tb!]
	\centering
	\small
		\begin{tabular}{|c|c||c|c|}
			\hline
			{\bf $W$} & {\bf $S$} & {\bf MCOD in \cite{TranFS16}}  & {\bf pMCOD} \\ \hline	\hline
			10K & 5\% & 100 (approx.) &	15.042 \\ \hline
			100K & 5\% & 700 (approx.) &	160.44 \\ \hline			
			100K & 10\% & 1420 (approx.) &	335.5 \\ \hline	
			100K & 20\% & 2650 (approx.)&	771.15 \\ \hline	
			100K & 50\% & 5270 (approx.) &	1981.25 \\ \hline										
		\end{tabular}
\caption{Average time to process a slide  in different works (in msecs).}
\label{table:comparison}
\end{table}

Finally, we compare our results regarding the Stock dataset against those of non-parallel \emph{MCOD} \cite{KontakiGPTM16}, as  evaluated by third parties in  \cite{TranFS16}.
The evaluation in \cite{TranFS16} also uses a processor  with clock speed at 3.5GHz, but without giving details about the number of processors. Nevertheless, our results can directly compare against those in Figures 6 and 10 in \cite{TranFS16}. Due to the log scale used, we can only report approximate values, as shown in Table \ref{table:comparison}. The speedup is between 2.66X and 6.65X, which provides strong insights into the parallelization efficiency of our solutions on a 4-core machine.

\subsubsection{pMCOD's features}
\label{sec:pmcod-features}

\begin{figure}[tb!]
	\centerline{\includegraphics[width=0.65\linewidth]{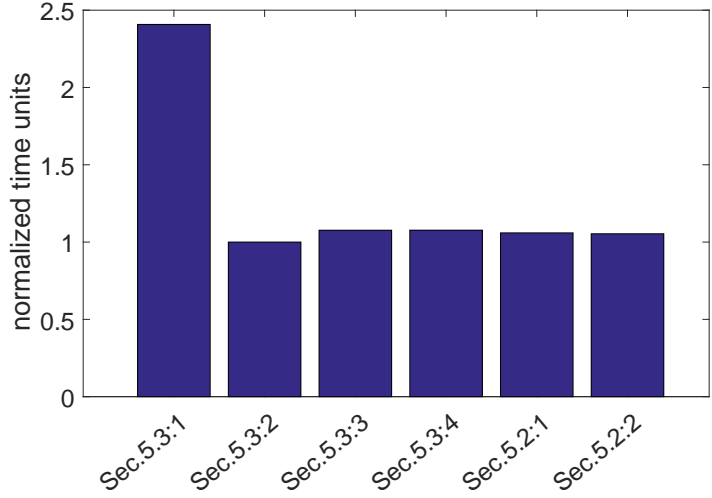}}
	\caption{Normalized times over all datasets, and window and slide sizes for the alternatives of pMCOD in Section \ref{sec:pmcod}.}
	\label{fig:pmcod_features}
\end{figure}

Thus far, \emph{pMCOD} has been the best-performing technique using the rationale in Section \ref{sec:pmcod-mc} (first alternative in Section \ref{subsec:mtree}). To assess the effectiveness of the event queue and the more restricted usage of M-tree, we run experiments over all datasets and window sizes and slides employed previously keeping the portion of outliers approximately at 1\%. The normalized results appear in Figure \ref{fig:pmcod_features}. From the figure, the following observations can be extracted:
\begin{enumerate}
	\item Employing the M-tree for all data (as in our default mode) incurs a significant overhead. Micro-clustering is so effective in our experiments that renders any additional structure that has shown to be advantageous in other settings negligible. If we restrict the usage of the M-tree, as in the third and fourth flavors in Section \ref{subsec:mtree}, the overhead due to the need of continuous updates observed is 7.7\%.

	\item The usage of the event queue (see the two rightmost bars) incurs overhead that is not outweighed by the benefits. On average, the algorithm is slowed down by 5-6\% compared to the case without any additional structure. Also, the implementation using  a  \emph{Treeset} outperforms almost consistently  the one using \emph{Priority Queue} but by a negligible margin.
\end{enumerate}

The above observations show that the \emph{pMCOD} algorithm runs faster without an event queue and the usage of M-trees for the range queries. The main reason  lies in the use of micro-clusters that manage to cover almost all the non-outlier objects. In other words, not only the outliers are a very small fraction of the data, but the $\mathbb{PO}$ set is small as well. As such, the Occam's razor principle applies, according to which, between models of similar performance we should adopt the simpler one \cite{TanSK2005}. In the remainder of our experiments, the second and simpler flavor of \emph{pMCOD} that does not employ any additional data structures (see Section \ref{subsec:mtree}) is used.

\subsection{Experiments in a cluster}

The purpose of the experiments using a cluster is to provide better evidence about the behavior of the partitioning alternatives and the scalability of the approach, especially regarding the slide size and the data dimensionality, where the scalability using a single machine has been problematic. Also, we strengthen the experimental evidence that \emph{pMCOD} is the best-performing technique.

The cluster consists of three machines with similar characteristics in terms of CPU speed. Apart from the 4-core/ 8-thread machine with 32 GB of RAM that we used in the previous experiments, we connect through 1 Gbps Ethernet a 6-core/12-thread machine with 64GB of RAM and another one with 8-core/8-threads and 32 GB of RAM.
The setup is as a standalone cluster with one \emph{Job Manager} on the 6-core/12-thread machine and one \emph{Task Manager} on each machine, i.e., overall we have three Task Managers.

\begin{figure}[tb!]
	\centerline{\includegraphics[width=0.65\linewidth]{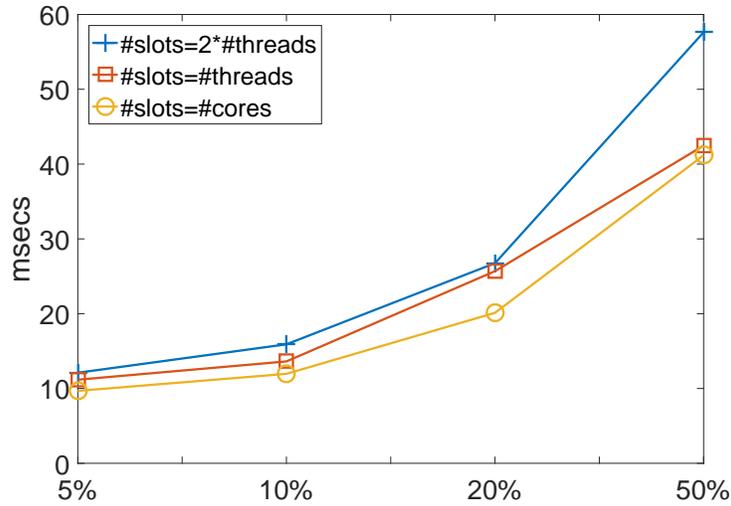}}
	\caption{Average processing time per slide for a 10K window with different numbers of Flink task slots.}
	\label{fig:task_slots}
\end{figure}

Before proceeding to the main experiments, we perform some basic tuning, since the results in Figure \ref{fig:pMCOD}(left) do not hold anymore. We employ the Stock dataset and a window of 10K objects, with the slide varying between 5\% and 50\% using the \emph{grid partitioning}, in order to gain some insight into the performance of the cluster when the number of the task slots varies. Figure \ref{fig:task_slots} shows that with the number of the task slots being equal to the number of the total cores in the cluster, the running time is on average 25\% lower than when the task slots are twice the number of threads and 12\% lower than when the task slots match the threads. Therefore, we fix the number of tasks accordingly.

\begin{figure}[tb!]
	\centerline{\includegraphics[width=0.65\linewidth]{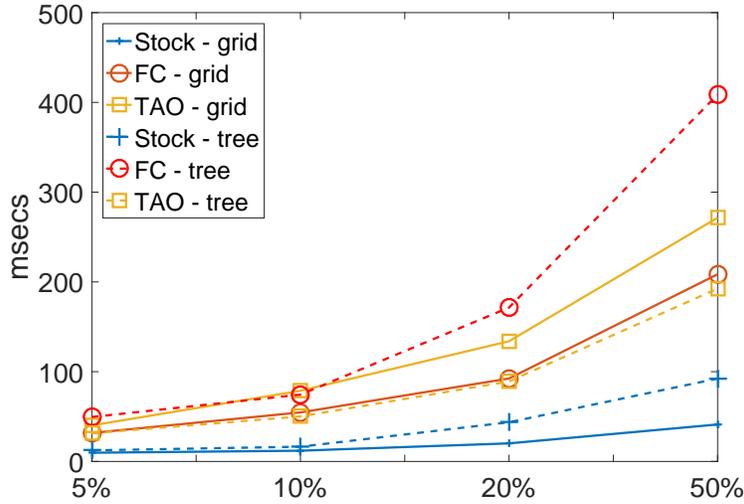}}
	\caption{Average processing time per slide for a 10K window with grid based and VP-tree based partitioning.}
	\label{fig:grid_vs_tree}
\end{figure}

\subsubsection{Partitioning and Slicing evaluation}

In the next experiment, we use the \emph{pMCOD} algorithm on the Stock (1-dimensional), FC (2-dimensional) and TAO (3-dimensional) datasets in order to compare the efficiency of the \emph{grid based} and the \emph{VP-tree based} partitioning. Each grid was created from statistics extracted from the whole dataset, while each VP-tree was created from the first 10K data points of each dataset. From  Figure \ref{fig:grid_vs_tree}, the following observations can be extracted:
\begin{enumerate}
    \item The observed speed-ups are much higher than previously. For instance, for the Stock dataset, when the slide size is 50\%, \emph{pMCOD} is 510X faster than the naive parallel solution.
    \item The scalability in the slide size is much better compared to the single-machine cases (we further elaborate on this in the next sub-section).
	\item In the 1 and 2-dimensional datasets, the grid-based partitioning is faster than the VP-tree-based one by 1.3 to 3 times. This stems from the fact that the replication rate of the VP-tree partitioning is higher than the replication rate of the grid.
	\item In the 3-dimensional dataset, the VP-tree partitioning is faster than the grid partitioning by approximately 1.4 times. This is due to the fact that its replication rate is lower than the rate from the grid partitioning. 
	\item Using the VP-tree based partitioning, the 3-dimen-sional dataset times are faster than those for the 2-dimensional dataset in most of the cases, which is due to both the lower replication rate in the 3-dimensional dataset and the difference in the total percentage of the outliers in the two datasets.
    \item  \emph{pMCOD} scales better with regards to the data dimensionality when the tree-based partitioning is used. For example, the ratios of the 3-dimensional dataset times to those of the single-dimensional for slides 5\%, 10\%, 20\% and 50\% are 2.5, 3.1, 2 and 2.1, respectively.   If, for the single-dimensional dataset, we consider the times of the grid-based partitioning that  are lower, the ratios become 3.2, 4.2, 4.4, and 4.7, respectively, which still denote much improved scalability.
\end{enumerate}

\begin{figure}[tb!]
	\centerline{\includegraphics[width=0.65\linewidth]{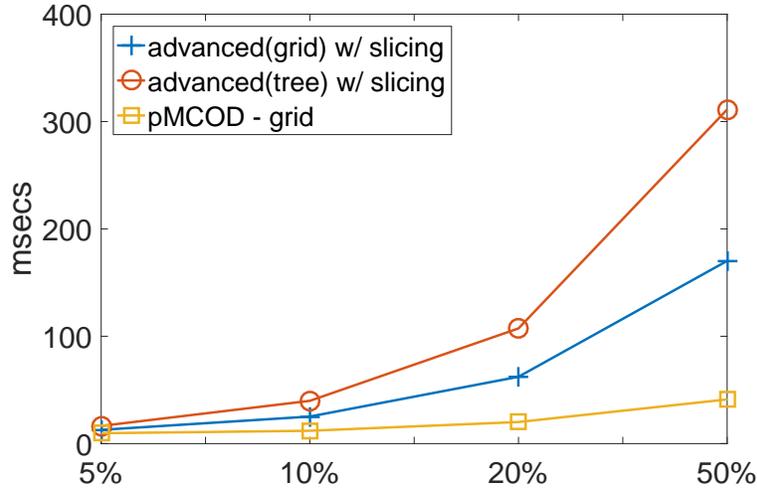}}
	\caption{Average  slide processing time for slide sizes of 5\%-50\% of \emph{pMCOD} and slicing-endowed advanced algorithms.}
	\label{fig:slicing}
\end{figure}

In order to extend the evidence that \emph{pMCOD} is the best-performing technique, we further compare  it against the flavors of the advanced solutions that follow the time-slicing rationale. Figure \ref{fig:slicing} shows this comparison for the Stock dataset. The \emph{pMCOD } algorithm remains the clear winner. Also, as previously, the grid partitioning outperforms the tree-based one for the Stock dataset for the advanced algorithm, too.

\begin{figure}[tb!]
	\centerline{\includegraphics[width=0.65\linewidth]{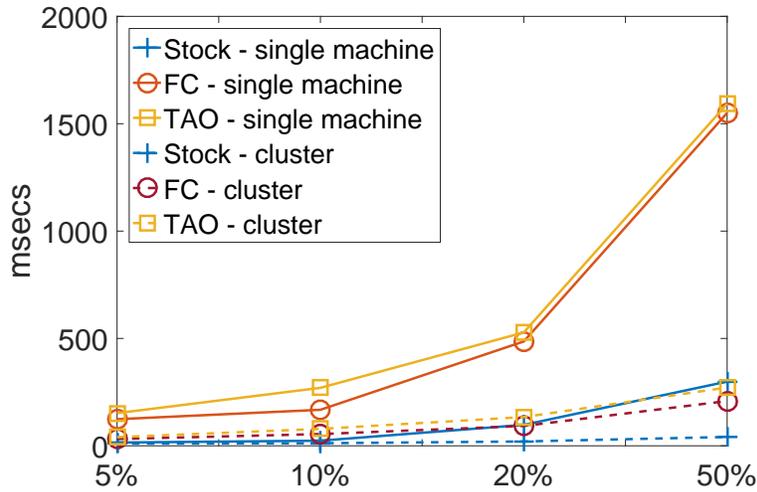}}
	\caption{Average  slide processing time for slide sizes of 5\%-50\% in the two main experimental settings.}
	\label{fig:cluster_vs_single}
\end{figure}

\subsubsection{Further experiments regarding the scalability}
With the help of Figure \ref{fig:cluster_vs_single}, we continue the discussion about the improved scalability regarding the slide size in a cluster setting.

First, when the slide size doubles, the running time increases by a smaller margin in the cluster.

Further, we can observe that, for large slides (e.g., 50\%), the speed-up is super-linear. For example, the times in the cluster for the 1, 2 and 3-dimensional datasets are 7.3, 7.4 and 5.9 times lower than the times on a single machine, respectively. Since the single machine possesses 4 cores and the cluster consists of 18 cores, a linear speed-up corresponds to 4.5 times lower. Superlinear speed-ups are not uncommon in MapReduce-like environments, due to the reduced contention of the system as the number of machines increases \cite{GPF15}.


Finally, we have already argued about the improved scalability in data dimensionality up to 3 dimensions. In
another experiment, we used a new 5-dimensional dataset by selecting the first 5 dimensions of the FC dataset and normalizing the values. With $R=0.182$ and $k=50$ approximately 1.2\% data points of each window are outliers while the partitioning is done using the VP-tree based technique. In such a case, the average processing time per slide size is, on average, more than 50 times higher than the corresponding times for Stock; compared to the 2-d FC dataset, the
increases are over 12X. This implies that many issues remain to be resolved for efficient parallel outlier mining in high-dimensional data.

\section{Related Work}
\label{sec:rw}

Outlier detection has been a topic that has attracted a lot of interest and there are several comprehensive surveys, e.g., \cite{GGAH14,ChandolaBK12,AkogluTK15,Aggarwal2013}.
In Section \ref{sec:back}, we have already discussed algorithms for outlier detection in streams. The next most related area to our work is parallel algorithms for distance-based outlier detection. A distributed outlier detection algorithm for massive datasets is proposed in \cite{CYK+17}. The two key points of this research is the initial partitioning of the data and the different outlier detection algorithm that each partition may run. The partitioning resembles our value-based one and focuses on the workload that each partition receives. Then, in each partition, the exact outlier detection algorithm out of two candidates is chosen. However, none of these algorithms is suitable in a streaming setting. There are also some works that assume parallel infrastructures that cannot scale and do not follow the paradigm introduced by MapReduce and its modern extensions, e.g., \cite{AngiulliBLS13,BhaduriMG11}. Overall, our work is the first one that combines streaming and massively parallel solutions to the problem of distance-based outlier detection. However, parallel and streaming anomaly detection has been considered for other definitions of outlierness, e.g., as in \cite{RettigKCP15}.

A related yet different problem is examined in  \cite{yan2015distributed}.
In a production distributed environment, a stream of data points may be split across multiple nodes, each holding
part of the values of a data point.
These parts will eventually need to be aggregated on a core node for outlier detection, but this incurs significant communication cost.
The solution proposed is based on compressing local data into a sketch.
Another related problem is that of supporting multiple outlier detection queries, i.e., combinations of $R$ and $k$ values. Examples include \cite{cao2016sharing} and \cite{KontakiGPTM16}. The latter presents multi-query extensions to MCOD and its approach is compatible with the our parallel pMCOD solution; here, we have examined single-query solutions only.

Finally, apart from the platforms discussed, there are additional alternatives.
For example, ChronoStream \cite{wu2015chronostream} is a prototype system for elastic big stream processing in a distributed environment, providing low latency.
However, we have decided to adopt Flink because it combines strong positive features, as discussed in Section \ref{sec:frameworks}, with mature engineering and a broad user community, while we did not consider elasticity issues in this work.


\section{Concluding Remarks}
\label{sec:disc}

This work targets streaming distance-based outlier detection and provides the first solutions to date to this problem, when examined in a massively parallel setting, such as Flink.
We have proposed a series of alternative techniques,
with the one termed as \emph{pMCOD} being a clear winner in the experiments that we have conducted using three real-world and one synthetic dataset. The improvements upon other solutions are significant, if not impressive, reaching up to an order of magnitude compared to the second best solution and up to three orders of magnitude compared to baseline solutions. There are also good speedups, between 2.66 and 6.65 times, compared to the non-parallel solutions implemented by third parties in \cite{TranFS16}, when running on a 4-core machine. Even higher performance is achieved in a small cluster.
Also, our solutions have been made publicly available.
The motivation behind our work is to fill a gap in the currently offered solutions in large-scale streaming big data analytics.
Moreover, our solutions aspire to act as a reference point for
future techniques that target both continuous reporting of distance-based outliers and a massively parallel setting; to this end, the alternative techniques are not tailored to Flink but they can be transferred to other similar frameworks.

We identify three avenues for such future extensions.
First, further research is required in order to make VP-tree based partitioning more practical and adaptively balanced, possibly using different techniques for the initial construction of the tree, addressing also the issue of acquiring both initial and evolving metadata to reach efficient partitioning decisions. Second, the current solutions, similarly to any distance-based outlier detection technique, are sensitive to the input parameters regarding the neighborhood radius and the threshold on the neighbors; evaluating multiple parameters in parallel is a promising approach to tackle this drawback, but, to this end, several issues need to be resolved. Third, further improvements need to be made in order to address the high-dimensionality problems of data possibly adopting additional definitions of outlierness and thus departing from the distance-based one and leveraging subspace techniques. Finally, another line of future research relates to approximate outlier mining; here, we have provided exact solutions only.


\begin{thebibliography}{10}
\expandafter\ifx\csname url\endcsname\relax
  \def\url#1{\texttt{#1}}\fi
\expandafter\ifx\csname urlprefix\endcsname\relax\def\urlprefix{URL }\fi
\expandafter\ifx\csname href\endcsname\relax
  \def\href#1#2{#2} \def\path#1{#1}\fi

\bibitem{Aggarwal2013}
C.~C. Aggarwal, Outlier Analysis, 2ed, Springer, 2018 (2018).

\bibitem{Johnson92}
R.~Johnson, Applied Multivariate Statistical Analysis, Prentice Hall, 1992
  (1992).

\bibitem{KNT00}
E.~Knorr, R.~Ng, V.~Tucakov, Distance-based outliers: algorithms and
  applications, The VLDB Journal 8~(3-4) (2000) 237--253 (2000).

\bibitem{KontakiGPTM16}
M.~Kontaki, A.~Gounaris, A.~N. Papadopoulos, K.~Tsichlas, Y.~Manolopoulos,
  Efficient and flexible algorithms for monitoring distance-based outliers over
  data streams, Inf. Syst. 55 (2016) 37--53 (2016).

\bibitem{YRW09}
D.~Yang, E.~Rundensteiner, M.~Ward, Neighbor-based pattern detection for
  windows over streaming data, in: EDBT, 2009, pp. 529--540 (2009).

\bibitem{cao2014scalable}
L.~Cao, D.~Yang, Q.~Wang, Y.~Yu, J.~Wang, E.~A. Rundensteiner, Scalable
  distance-based outlier detection over high-volume data streams, in: {ICDE},
  2014, pp. 76--87 (2014).

\bibitem{AF07}
F.~Angiulli, F.~Fassetti, Detecting distance-based outliers in streams of data,
  in: CIKM, 2007, pp. 811--820 (2007).

\bibitem{CYK+17}
L.~Cao, Y.~Yan, C.~Kuhlman, Q.~Wang, E.~A. Rundensteiner, M.~Y. Eltabakh,
  Multi-tactic distance-based outlier detection, in: {ICDE}, 2017, pp. 959--970
  (2017).

\bibitem{TranFS16}
L.~Tran, L.~Fan, C.~Shahabi, Distance-based outlier detection in data streams,
  {PVLDB} 9~(12) (2016) 1089--1100 (2016).

\bibitem{DSAA18}
T.~Toliopoulos, A.~Gounaris, K.~Tsichlas, A.~Papadopoulos, S.~Sampaio, Parallel
  continuous outlier mining~in streaming data, in: 5th {IEEE} International
  Conference on Data Science and Advanced Analytics (DSAA), 2018 (2018).

\bibitem{KGP+11}
M.~Kontaki, A.~Gounaris, A.~N. Papadopoulos, K.~Tsichlas, Y.~Manolopoulos,
  Continuous monitoring of distance-based outliers over data streams, in: ICDE,
  2011, pp. 135--146 (2011).

\bibitem{KGP+13}
M.~Kontaki, A.~Gounaris, A.~N. Papadopoulos, K.~Tsichlas, Y.~Manolopoulos,
  Continuous outlier detection in data streams: an extensible framework and
  state-of-the-art algorithms, in: SIGMOD, 2013, pp. 1061--1064 (2013).

\bibitem{WR09}
S.~Wang, E.~A. Rundensteiner, Scalable stream join processing with expensive
  predicates: workload distribution and adaptation by time-slicing, in: EDBT,
  2009, pp. 299--310 (2009).

\bibitem{CPZ97}
P.~Ciaccia, M.~Patella, P.~Zezula, M-tree: An efficient access method for
  similarity search in metric spaces, in: VLDB, 1997, pp. 426--435 (1997).

\bibitem{yianilos1993data}
P.~N. Yianilos, Data structures and algorithms for nearest neighbor search in
  general metric spaces, in: SODA, Vol.~93, 1993, pp. 311--321 (1993).

\bibitem{TanSK2005}
P.~Tan, M.~Steinbach, V.~Kumar, Introduction to Data Mining, Addison-Wesley,
  2005 (2005).

\bibitem{GPF15}
N.~J. Gunther, P.~Puglia, K.~Tomasette, Hadoop superlinear scalability, Commun.
  ACM 58~(4) (2015) 46--55 (2015).

\bibitem{GGAH14}
M.~Gupta, J.~Gao, C.~C. Aggarwal, J.~Han, Outlier detection for temporal data:
  {A} survey, {IEEE} Trans. Knowl. Data Eng. 26~(9) (2014) 2250--2267 (2014).

\bibitem{ChandolaBK12}
V.~Chandola, A.~Banerjee, V.~Kumar, Anomaly detection for discrete sequences:
  {A} survey, {IEEE} Trans. Knowl. Data Eng. 24~(5) (2012) 823--839 (2012).

\bibitem{AkogluTK15}
L.~Akoglu, H.~Tong, D.~Koutra, Graph based anomaly detection and description: a
  survey, Data Min. Knowl. Discov. 29~(3) (2015) 626--688 (2015).

\bibitem{AngiulliBLS13}
F.~Angiulli, S.~Basta, S.~Lodi, C.~Sartori, Distributed strategies for mining
  outliers in large data sets, {IEEE} Trans. Knowl. Data Eng. 25~(7) (2013)
  1520--1532 (2013).

\bibitem{BhaduriMG11}
K.~Bhaduri, B.~L. Matthews, C.~Giannella, Algorithms for speeding up
  distance-based outlier detection, in: Proc. of {SIGKDD}, 2011, pp. 859--867
  (2011).

\bibitem{RettigKCP15}
L.~Rettig, M.~Khayati, P.~Cudr{\'{e}}{-}Mauroux, M.~Pi{\'{o}}rkowski, Online
  anomaly detection over big data streams, in: Big Data, 2015, pp. 1113--1122
  (2015).

\bibitem{yan2015distributed}
Y.~Yan, J.~Zhang, B.~Huang, X.~Sun, J.~Mu, Z.~Zhang, T.~Moscibroda, Distributed
  outlier detection using compressive sensing, in: Proc. of {SIGMOD}, ACM,
  2015, pp. 3--16 (2015).

\bibitem{cao2016sharing}
L.~Cao, J.~Wang, E.~A. Rundensteiner, Sharing-aware outlier analytics over
  high-volume data streams, in: Proc. of {SIGMOD}, 2016, pp. 527--540 (2016).

\bibitem{wu2015chronostream}
Y.~Wu, K.-L. Tan, Chronostream: Elastic stateful stream computation in the
  cloud, in: {ICDE}, IEEE, 2015, pp. 723--734 (2015).

\end{thebibliography}

\end{document}